\documentclass[twocolumn]{aastex631}
\usepackage{graphicx}
\usepackage{epstopdf}
\usepackage{subfigure} 
\usepackage{color}
\usepackage{float}
\usepackage[figuresright]{rotating}
\usepackage{footnote}
\usepackage{amsmath}
\usepackage{subfigure} 
\usepackage{appendix}
\usepackage{longtable}
\usepackage{threeparttable}
\usepackage{booktabs}
\usepackage{hyperref}
\usepackage{CJKutf8}

\newcommand{\teff}{$T_{\rm eff}$}
\newcommand{\logg}{$\log g$}

\begin{document}
\begin{CJK}{UTF8}{gbsn}
\title{Discoveries and Properties of EL CVn-type Binaries in the TESS Survey}

\author[0000-0003-4829-6245]{Jianping Xiong}
\affiliation{Yunnan Observatories, Chinese Academy of Sciences, 396 YangFangWang, Guandu District, Kunming, 650216, Peopleʼs Republic of China}
\affiliation{Key Laboratory for Structure and Evolution of Celestial Objects, Chinese Academy of Sciences, P.O. Box 110, Kunming 650216, People's Republic of China}
\affiliation{International Centre of Supernovae, Yunnan Key Laboratory, Kunming 650216, People's Republic of China}

\author[0000-0002-1421-4427]{Zhenwei Li}
\affiliation{Yunnan Observatories, Chinese Academy of Sciences, 396 YangFangWang, Guandu District, Kunming, 650216, Peopleʼs Republic of China}
\affiliation{Key Laboratory for Structure and Evolution of Celestial Objects, Chinese Academy of Sciences, P.O. Box 110, Kunming 650216, People's Republic of China}
\affiliation{International Centre of Supernovae, Yunnan Key Laboratory, Kunming 650216, People's Republic of China}

\author[0000-0002-2577-1990]{Jiao Li}
\affiliation{Yunnan Observatories, Chinese Academy of Sciences, 396 YangFangWang, Guandu District, Kunming, 650216, Peopleʼs Republic of China}
\affiliation{Key Laboratory for Structure and Evolution of Celestial Objects, Chinese Academy of Sciences, P.O. Box 110, Kunming 650216, People's Republic of China}
\affiliation{International Centre of Supernovae, Yunnan Key Laboratory, Kunming 650216, People's Republic of China}

\author{Xiaobin Zhang}
\affiliation{CAS Key Laboratory of Optical Astronomy, National Astronomical Observatories, Chinese Academy of Sciences,\\ Beijing 100101, China}
\affiliation{School of Astronomy and Space Science, University of the Chinese Academy of Sciences, Beijing 101408, China} 
\affiliation{Department of Astronomy, China West Normal University, Nanchong, China}

\author[0000-0001-7084-0484]{Xiaodian Chen}
\affiliation{CAS Key Laboratory of Optical Astronomy, National Astronomical Observatories, Chinese Academy of Sciences,\\ Beijing 100101, China}
\affiliation{School of Astronomy and Space Science, University of the Chinese Academy of Sciences, Beijing 101408, China} 
\affiliation{Department of Astronomy, China West Normal University, Nanchong, China}

\author{Kaifan Ji}
\affiliation{Yunnan Observatories, Chinese Academy of Sciences, 396 YangFangWang, Guandu District, Kunming, 650216, Peopleʼs Republic of China}

\author[0000-0001-9204-7778]{Zhanwen Han}
\affiliation{Yunnan Observatories, Chinese Academy of Sciences, 396 YangFangWang, Guandu District, Kunming, 650216, Peopleʼs Republic of China}
\affiliation{Key Laboratory for Structure and Evolution of Celestial Objects, Chinese Academy of Sciences, P.O. Box 110, Kunming 650216, People's Republic of China}
\affiliation{International Centre of Supernovae, Yunnan Key Laboratory, Kunming 650216, People's Republic of China}

\author[0000-0001-5284-8001]{Xuefei Chen}
\affiliation{Yunnan Observatories, Chinese Academy of Sciences, 396 YangFangWang, Guandu District, Kunming, 650216, Peopleʼs Republic of China}
\affiliation{Key Laboratory for Structure and Evolution of Celestial Objects, Chinese Academy of Sciences, P.O. Box 110, Kunming 650216, People's Republic of China}
\affiliation{International Centre of Supernovae, Yunnan Key Laboratory, Kunming 650216, People's Republic of China}
\correspondingauthor{Jianping Xiong, Xuefei Chen}
\email{xiongjianping@ynao.ac.cn, cxf@ynao.ac.cn}



\begin{abstract}

EL CVn-type systems represent a rare evolutionary stage in binary star evolution, providing ideal laboratories for investigating stable mass transfer processes and the formation of extremely low-mass white dwarfs (ELM WDs). The Transiting Exoplanet Survey Satellite (TESS) has delivered an extensive collection of high-precision time-domain photometric data, which is invaluable for studying EL CVn binaries. In this study, we identified 29 EL CVn systems from the TESS eclipsing binary catalogs (sectors 1–65), 11 of which are newly discovered. These systems consist of smaller, hotter pre-He white dwarfs and A/F main-sequence stars. The orbital periods of these binaries range from 0.64 to 2.5 days. Utilizing TESS light curves, Gaia distances, and multi-band photometric data (e.g., GALEX, 2MASS, WISE, SkyMapper), we modeled the light curves and spectral energy distributions to derive system parameters, including effective temperatures, masses, and radii. These systems were then compared with the white dwarf mass-period relation and the evolutionary tracks of ELM WDs. The comparison reveals that these binaries are consistent with the expected mass-period relation for white dwarfs and align well with the evolutionary tracks on the \teff\,-\logg\, diagram for ELM WDs. This result suggests that these EL CVn systems likely formed through stable mass transfer processes. We provide a catalog of complete parameters for 29 EL CVn systems identified from the TESS survey. This catalog will serve as an essential resource for studying binary mass transfer, white dwarf formation, and pulsation phenomena in EL CVn-type systems.

\end{abstract}



\section{Introduction} \label{sec:intro}
EL CVn-type binaries are a distinctive class of eclipsing binary systems, comprising an A/F-type or B-type main-sequence (MS) star and a low-mass ($\sim0.15$-$0.33\rm{M}_{\odot}$) pre-helium white dwarf (pre-He WD) \citep{2011MNRAS.418.1156M,2014MNRAS.437.1681M}. These systems are the post-mass transfer remnants of close binary evolution, typically formed through stable mass transfer initiated by Roche-lobe overflow (RLOF) \citep{2017MNRAS.467.1874C}. During this process, the progenitor of the pre-helium white dwarf, originally the more massive star, evolves into a red giant and transfers its outer envelope to its companion, leaving behind the exposed core of a stripped red giant \citep{2010ApJ...715...51V, 2011MNRAS.418.1156M}. This evolutionary pathway not only sheds light on the terminal phases of low-mass star evolution, but also offers a valuable laboratory for studying the dynamics of mass transfer in close binary systems.

The EL CVn-type binaries with smaller and hotter pre-He WD secondaries exhibit distinct features in time-domain photometric observations. In the folded light curve, such systems are characterized by a deeper secondary eclipse, where the pre-He WD is eclipsed by the MS star, displaying a "boxy" shape with steep ingress and egress phases and a flat-bottomed minimum. In contrast, the primary eclipse, caused by the transit of the pre-He WD across the MS star, is notably shallower.
Recent photometric surveys have significantly expanded the number of known EL CVn-type binaries. And approximately 80 such systems were detected in the WASP (Wide Angle Search for Planets, \citealt{2011MNRAS.418.1156M,2014MNRAS.437.1681M,2020NewA...7801363L, 2020AJ....160...49L,2021AJ....161..137H,2021AJ....162..212K,2022MNRAS.511..654L,2022MNRAS.515.4702L}), Kepler \citep{2010ApJ...715...51V,2012ApJ...748..115B,2011ApJ...728..139C,2012ApJ...748..115B,2015ApJ...803...82R,2017ApJ...837..114G,2017ApJ...850..125Z,2019ApJ...884..165Z}, PTF (Palomar Transient Factory, \citet{2018MNRAS.475.2560V}), TESS (Transiting Exoplanet Survey Satellite, \citealt{2020ApJ...888...49W,2024NewA..10702153P,2024MNRAS.533.2058C}) and Gaia \citep{2023A&A...674A..34G}. In addition, for EL CVn systems with smaller and cooler pre-He WD secondaries, the beaming, ellipsoidal, and reflection effects are observed along with primary and secondary eclipses. These systems have also been identified in the Kepler survey \citep{2015ApJ...815...26F}.

To rigorously test theoretical models of white dwarf cooling and binary evolution, precise measurements of the effective temperatures, luminosities, masses, and radii of EL CVn-type binaries are essential. Among the discovered $\sim$80 systems, only about half have had their full set of parameters, including masses and radii, determined. The further analysis of some systems has been hindered by the lack of complete light curves, limiting our ability to derive critical parameters. Moreover, only a limited number of these systems have been subjected to multi-epoch spectroscopic observations \citep{2011MNRAS.418.1156M,2020AJ....159....4W,2024MNRAS.533.2058C}, which are essential for obtaining accurate dynamical masses and refining our understanding of their evolutionary pathways.

Since 2018, the TESS survey has provided high-precision, continuous, and long-duration photometric observations, enabling the study of binary systems across nearly the entire sky. To date, more than 10,000 eclipsing binaries have been identified using TESS data \citep{2021AA...652A.120I,2022ApJS..258...16P,2022RNAAS...6...96H,2022ApJS..263...34C} (Gao et al., in prep). Furthermore, data from Gaia \citep{2023A&A...674A...1G, 2023A&A...674A..37G,2023MNRAS.524.1855Z} and LAMOST have provided highly accurate astrometric and atmospheric parameters for millions of stars within the Galaxy. Additionally, large multi-band photometric surveys such as GALEX (Galaxy Evolution Explorer; \citealt{2014AdSpR..53..900B}), 2MASS, the Wide-field Infrared Survey Explorer (WISE; W1 and W2; \citealt{2010AJ....140.1868W}), SkyMapper \citep{2019PASA...36...33O}, SDSS \citep{2000AJ....120.1579Y, 2017ApJS..233...25A}, and APASS (AAVSO Photometric All-Sky Survey; \citealt{2014CoSka..43..518H}) have contributed additional photometric data across multiple wavelengths, enabling a comprehensive investigation of EL CVn-type binaries and more precise parameter determination.

In this paper, we first conduct a search for EL CVn-type systems with smaller and hotter pre-He WD secondaries in the TESS survey. We then combine the astrometric and atmospheric parameters from Gaia and LAMOST to derive the physical parameters of the identified EL CVn binaries. Finally, we analyze the properties of these systems and discuss the mechanisms driving their evolution. The structure of the paper is as follows: Section~\ref{sec:data} describes the reduction of the observational data from the TESS survey. Section~\ref{sec:identify} details the identification process for EL CVn-type binaries. In Section~\ref{sec:method}, we present the methods used to determine the parameters of the identified systems. Section~\ref{sec:discuss} discusses the properties and evolutionary pathways of the EL CVn binaries. Finally, Section~\ref{sec:summary} provides a summary of our findings.

\section{Data} \label{sec:data}
\subsection{Target selection}
In this paper, we begin with an initial sample of 11,164 identified eclipsing binaries (EBs) from the TESS survey, as presented by \citet{2021AA...652A.120I,2022ApJS..258...16P,2022ApJS..263...34C} and Gao et al. (in prep). Since EL CVn-type binaries are detached systems, we filter the sample based on the following strategy: \citep{2022ApJS..258...16P} provides the ``\texttt{Morph}" parameter to assess the morphology of light curves, and Gao et al. (in prep) classified the light curves into EA-type and EW-type based on their shapes. Consequently, we remove ellipsoidal light curves using the morphological parameter (\texttt{Morph} $> 0.7$; \citealt{2012AJ....143..123M}) and exclude EW-type light curves in Gao's catalog. For the remaining two catalogs of \citet{2021AA...652A.120I,2022ApJS..263...34C}, after removing duplicate sources, we included all the remaining targets. This process results in a refined sample of 9,854 eclipsing binaries. We then conduct a systematic search for EL CVn-type binaries within this selection.
\begin{figure*}[t]
  \centering
    \subfigure{
   \includegraphics[scale=0.4]{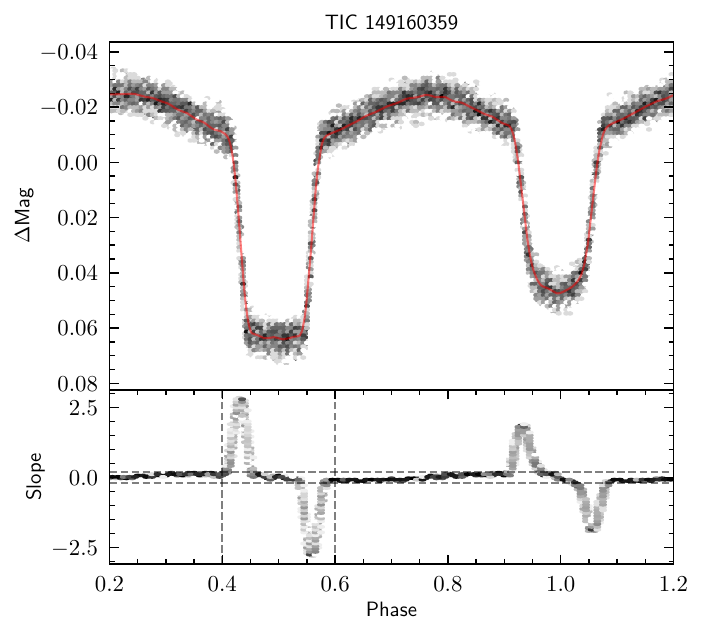}}
  \subfigure{
   \includegraphics[scale=0.4]{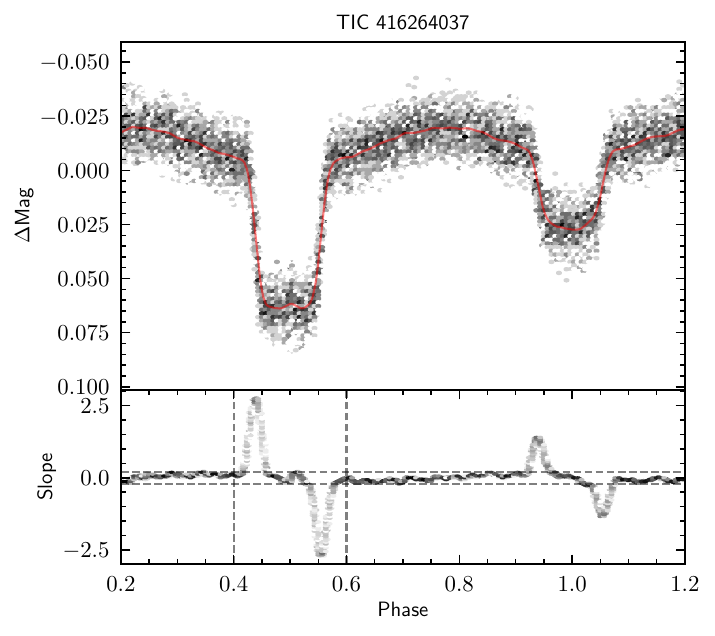}
  }
  \subfigure{
   \includegraphics[scale=0.4]{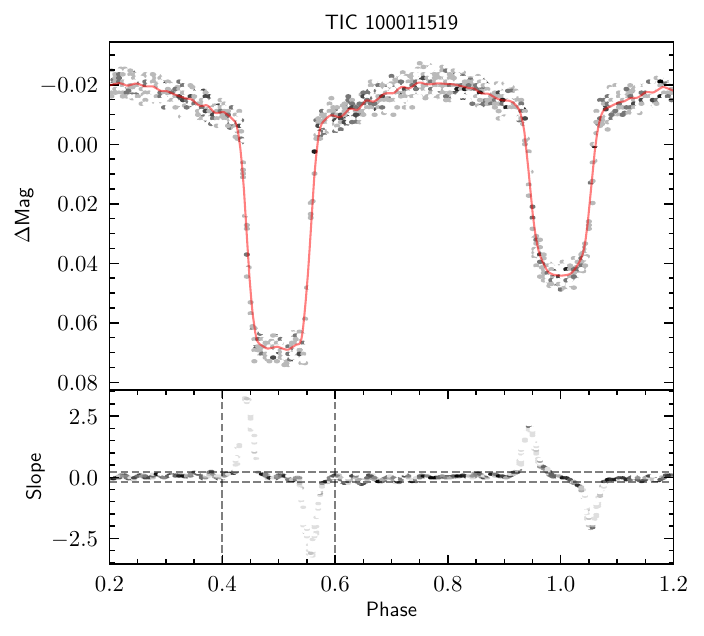}
  }
  
    \subfigure{
   \includegraphics[scale=0.4]{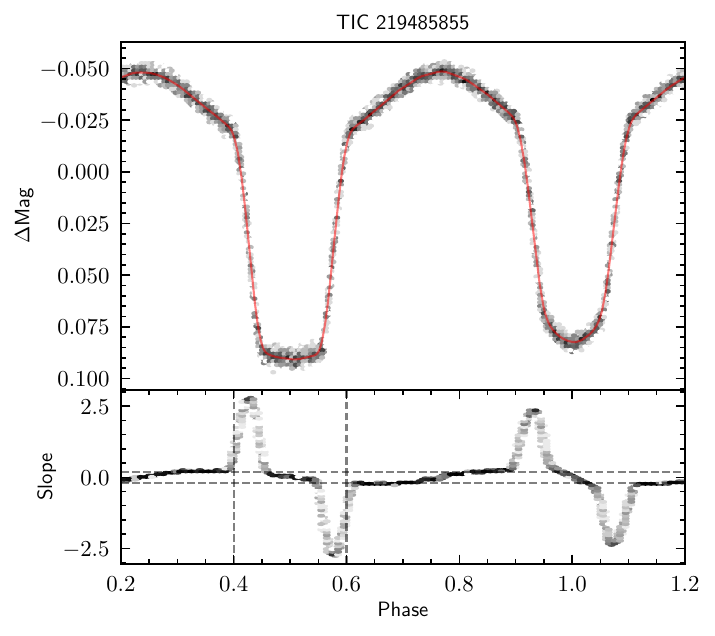}}
  \subfigure{
   \includegraphics[scale=0.4]{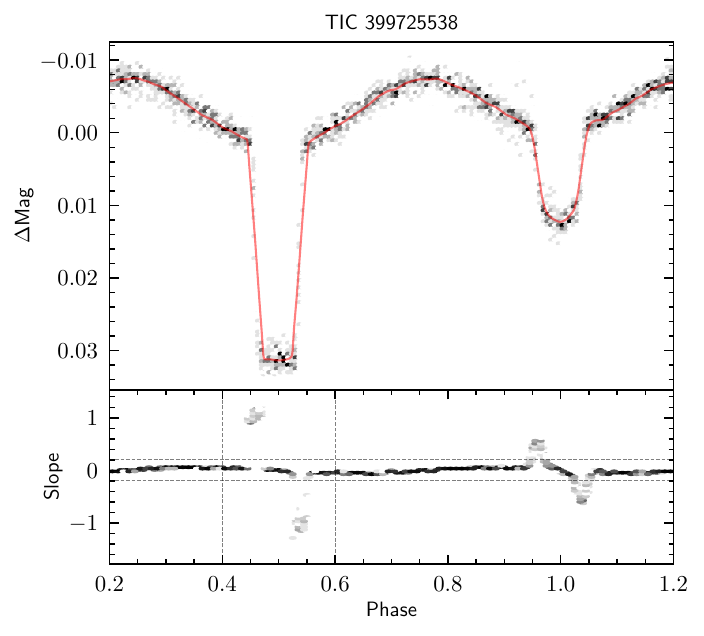}
  }
  \subfigure{
   \includegraphics[scale=0.4]{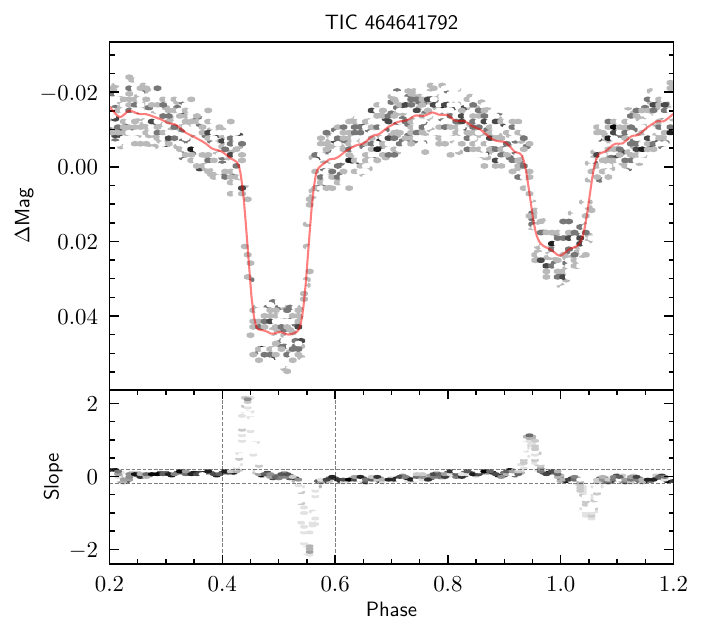}
  }
  \centering
  \caption{A schematic illustration of the slope calculations for six previously known EL CVn systems \citep{2020ApJ...888...49W,2024NewA..10702153P}. In the upper panel, the plot utilizes hexagonal bins (\texttt{"hexbin"}) to depict TESS observation data, where the shading of each bin corresponds to the density of data points within the respective region. Darker shades indicate a higher concentration of points within a hexagonal cell.
The red solid line shows the result of locally weighted scatterplot smoothing (LOWESS). In the lower panel, the slopes of the LOWESS-smoothed light curves are displayed, where the individual data points are similarly visualized using grayscale cells for improved clarity.}\label{fig:slope}
\end{figure*}
\subsection{Photometric Data}

The eclipsing binaries selected as initial samples were identified using short-cadence (2-minute) and long-cadence (30-minute) data from sectors 1-65 of the TESS survey. Therefore, we retrieve the TESS light curves for these binaries with both 2-minute and 30-minute cadence from the public data releases, processed by TESS-SPOC \footnote{DOI:\url{10.17909/t9-wpz1-8s54} \citep{2020RNAAS...4..201C}} (Science Processing Operations Center) and the MIT Quick Look Pipeline (QLP)\footnote{DOI:\url{10.17909/t9-r086-e880}\citep{2020RNAAS...4..204H,2020RNAAS...4..206H,2021RNAAS...5..234K,2022RNAAS...6..236K}}. The data are downloaded using cURL scripts\footnote{\url{https://archive.stsci.edu/tess/bulk_downloads/bulk_downloads_ffi-tp-lc-dv.html}} and Lightkurve module \citep{2018ascl}, provided by the Mikulski Archive for Space Telescopes (MAST\footnote{\url{https://archive.stsci.edu/missions-and-data/tess/data-products}}). 

We prioritize the use of 2-minute cadence data. If the 2-minute data are unavailable or invalid, we utilize the 30-minute cadence data as an alternative. Specifically, for the 2-minute cadence light curves, we use the \texttt{PDCSAP} flux \citep{2012PASP..124.1000S,2014PASP..126..100S}, while the 30-minute cadence data employ the \texttt{KSPSAP} flux \citep{2020RNAAS...4..204H,2020RNAAS...4..206H}. The observed light curves are folded according to the periods of the eclipsing binaries provided in the literature. Given that the defining feature of an EL CVn system is a deeper, ``boxy" secondary eclipse, the light curves are uniformly folded with the deeper eclipses aligned at phase=0.5, covering a phase range from 0.2 to 1.2. Before folding, we conduct a thorough quality control assessment, removing any data points flagged with a "quality" value $\neq$ 0 in the light curve, which indicates potential anomalies during the TESS pipeline processing. Subsequently, the light curves are normalized using equation~\ref{Equ:normalized}:
\begin{equation}\label{Equ:normalized}
\begin{aligned}
    m_{i}&=-2.5\times \mathrm{log\rm_{10}}(f_{i})\\
    m^{'}_{i}&=m_{i}-\frac{\sum^{n}_{i}m_{i}}{n}
    \end{aligned}
\end{equation}
In Eq.\ref{Equ:normalized}, $f\rm_{i}$ represents the observed light curve from TESS, $n$ denotes the number of light curves, and $m^{'}\rm_{i}$ is the normalized light curve expressed in the ``\texttt{mag}'' unit. 

\section{EL CVn detection} \label{sec:identify}
The EL CVn binaries are characterized by a deeper, ``boxy" secondary eclipse with a flat bottom. To identify this feature, we calculate the slope of the light curve. Due to the pulsations and the noise in the observation, at first, each original light curve is smoothed using Locally Weighted Scatterplot Smoothing (LOWESS), a technique commonly employed for data smoothing through locally weighted regression. During the smoothing process, 3\% of the data points are utilized for each local regression. Fig.\ref{fig:slope} employs 6 EL CVn systems that provided by \citet{2020ApJ...888...49W} and \citet{2024NewA..10702153P} to  illustrate the identification process. In the top panel of Fig.\ref{fig:slope}, the TESS observation data is represented using a grayscale plot, where the intensity of each grid cell corresponds to the number of data points within that region. This approach replaces the direct visualization of individual data points, allowing for a more intuitive interpretation of the overall data distribution. The red solid line illustrates the LOWESS-smoothed light curve. In the bottom panel, the slope calculations of the LOWESS-smoothed light curves are depicted, with individual points also visualized as grayscale cells to enhance clarity. The analysis then focuses on the phase range between 0.4 and 0.6, as marked by the two vertical dashed gray lines in the bottom panel. Next, we identify the two points corresponding to the maximum and minimum slopes between phases 0.4 and 0.6. The number of data points between these two points is then counted. If more than 10 data points exist in this interval, and over 50\% of them exhibit an absolute slope value of less than 0.2, the source is identified as a candidate EL CVn-type binary.
\begin{figure}[t]
  \centering
   \includegraphics[scale=0.7]{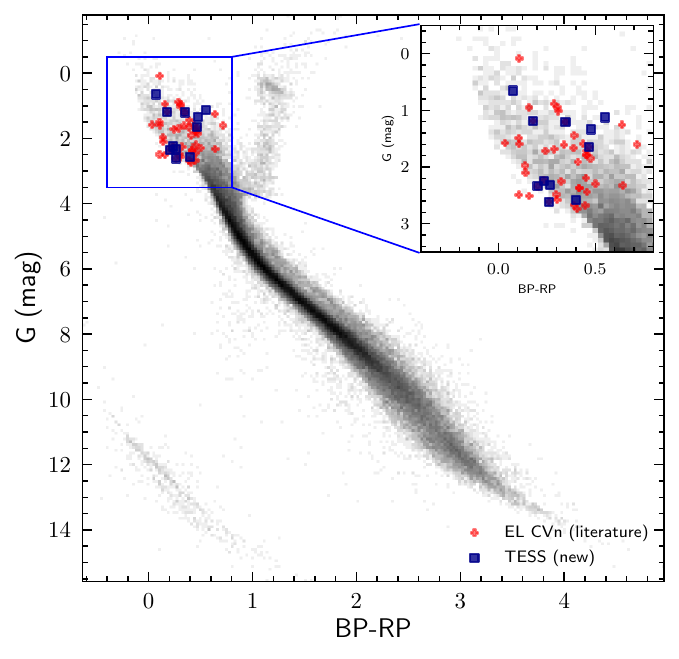}
  \centering
  \caption{The color-magnitude diagram of EL CVn systems. The grayscale plot represents the sources within 100 \texttt{pc} from Gaia DR3, with the intensity corresponding to stellar density. The red ``plus" markers indicate previously known EL CVn systems (see Sec. \ref{sec_intro}), while the newly discovered EL CVn candidates are represented by blue filled squares.}\label{fig:HRD}
\end{figure}

Subsequently, we apply additional selection criteria to further refine the sample. Specifically, we require the absolute difference between the depths of the two eclipses to be less than 0.15 \texttt{mag} and the eccentricity to be approximately zero \citep{2024NewA..10702153P}. Additionally, we visually inspect the light curves of the candidate systems, excluding those displaying EW-type features. Given that the primary optical contribution in EL CVn-type binaries comes from A/F main-sequence stars, we refer to the color-magnitude diagram in Fig.~\ref{fig:HRD}. 
In Fig.\ref{fig:HRD}, the background depicts sources within 100 \texttt{pc} from Gaia DR3. Similar to Fig.~\ref{fig:slope}, a grayscale plot is employed in place of individual points, with the intensity indicating stellar density within each grid region. The red plus sign markers indicate previously identified EL CVn systems (see Sec.\ref{sec_intro}). It is evident that most known EL CVn binaries are located on the A/F main sequence. To ensure consistency, we cross-match our candidates with Gaia DR3 using a 2 \texttt{arcsec} cone search and exclude stars with $e\_{\rm Plx}/\rm Plx > 0.2$. Additionally, we remove candidates that do not lie within the A/F main sequence region of the color-magnitude diagram. Consequently, we identify 29 EL CVn candidates from TESS, of which 18 were previously known, and 11 are newly discovered, as represented by blue filled squares in Fig.~\ref{fig:HRD}. Among these, 17 systems have yet to have their parameters determined. 
\begin{figure}[h]
  \centering
   \includegraphics[scale=0.75]{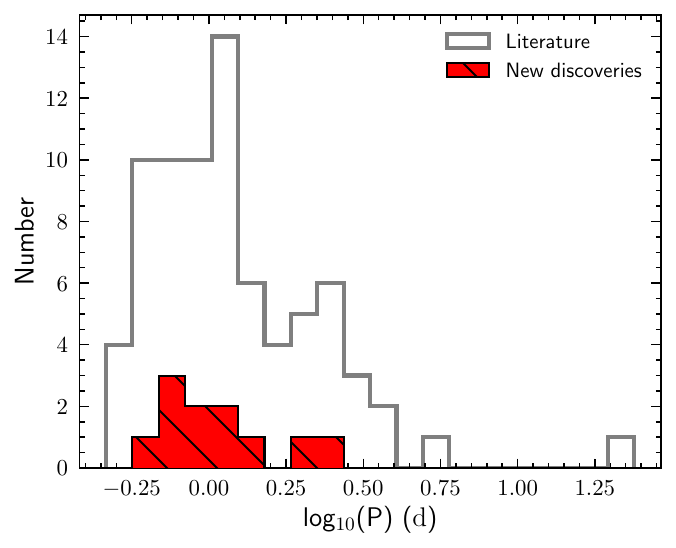}
  \centering
  \caption{The period distribution of EL CVn systems from observation. The gray histogram represents EL CVn binaries previously identified, while the red hatched histogram is the period distribution of the EL CVn binaries we discovered in TESS survey.}\label{fig:period}
\end{figure}

Fig.~\ref{fig:period} illustrates the period distribution of the discovered EL CVn systems. The gray histogram represents the EL CVn binaries previously identified in the literature (see Sect.\ref{sec_intro}), while the red hatched histogram indicates the period distribution of the EL CVn binaries discovered in the TESS survey. In Fig.~\ref{fig:period}, the orbital periods of previously discovered EL CVn-type systems range from 0.464 to 23.8776 \texttt{d}. And it can be seen that the periods of our newly discovered targets range from 0.64 to 2.5 \texttt{d}, thereby supplementing the previously sparse sample of those $>$ $\sim$1 \texttt{d}. Next, we will proceed to determine the parameters of these EL CVn systems identified from the TESS survey.

\section{Analysis} \label{sec:method}
\begin{table*}[t]
\caption{Comparison of the light curve fitting parameters of TIC 149160359 and TIC 416264037 between this work and \citet{2020ApJ...888...49W}.}\label{tab:table1}
\centering
\begin{tabular}{cccccc}
\hline
\hline
  \multicolumn{1}{c}{} &
  \multicolumn{2}{c}{TIC 149160359} &
  \multicolumn{2}{c}{TIC 416264037} \\
\hline

Parameters & \citet{2020ApJ...888...49W} & This work & \citet{2020ApJ...888...49W}  & This work\\
\hline
$i $ $(\circ)$ & 84.45 $\pm$ 0.04 & $89.66^{+0.02}_{-0.02}$ &80.31 $\pm$ 0.16  & $81.48^{+0.01}_{-0.01}$\\
\hline
$q $ & 0.0906 $\pm$ 0.0004 & $0.1002^{+0.0001}_{-0.0001}$ & 0.1086 $\pm$ 0.0027  & $0.1003^{+0.0001}_{-0.0001}$\\
\hline
$^{*}$ $r_{1} $ & 0.3900 $\pm$ 0.0002 & $0.3882^{+0.0000}_{-0.0000}$ & 0.3844 $\pm$ 0.0011  & $0.3850^{+0.0001}_{-0.0001}$\\
\hline
$^{*}$ $r_{2} $ & 0.0831 $\pm$ 0.0006 & $0.0873^{+0.0000}_{-0.0000}$ & 0.0683 $\pm$ 0.0028  & $0.0683^{+0.0000}_{-0.0000}$\\
\hline
$^{\bullet}$ $T_{2}/T_{1}$ & 1.134 $\pm$ 0.018 & $1.0966^{+0.0001}_{-0.0001}$ & 1.291 $\pm$0.017 & $1.3218^{+0.0002}_{-0.0002}$\\
\hline
\multicolumn{5}{l}{\footnotesize $^{*}$ The relative radii obtained from \citet{2020ApJ...888...49W} is $r_{pole}$}\\
\multicolumn{5}{l}{\footnotesize $^{\bullet}$ The effective temperature ratio for \citet{2020ApJ...888...49W} is derived from the absolute values.} \\ 
\multicolumn{5}{l}{  \footnotesize The uncertainty in the temperature ratio is propagated from the absolute uncertainties of}\\
\multicolumn{5}{l}{  \footnotesize the effective temperatures.}\\
\end{tabular}
\centering
\end{table*}

In principle, using the light curves and astrometric data provided by TESS and Gaia, along with multi-band photometry, we can derive the stellar parameters for the EL CVn candidate sources. To illustrate our parameter-solving process, we use TIC 149160359 and TIC 416264037 \citep{2020ApJ...888...49W} as examples.
\subsection{Light curve Fitting}
\begin{figure*}[t]
  \centering
    \subfigure{
   \includegraphics[scale=0.55]{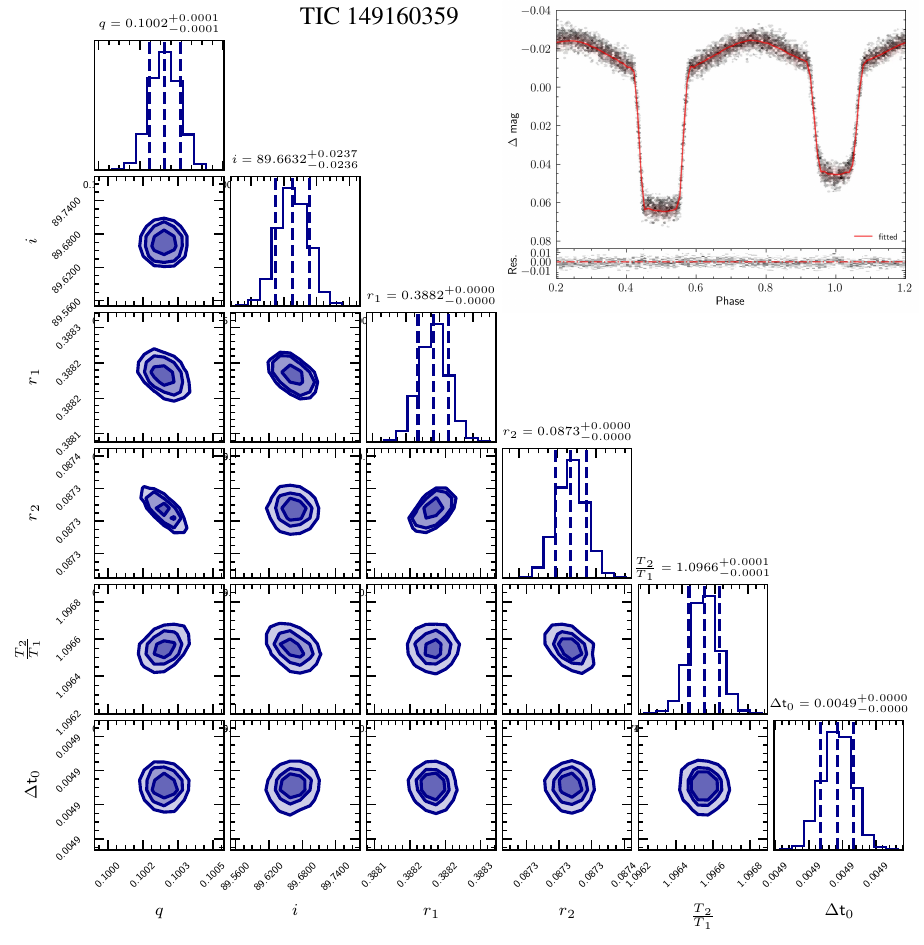}}
  \subfigure{
   \includegraphics[scale=0.55]{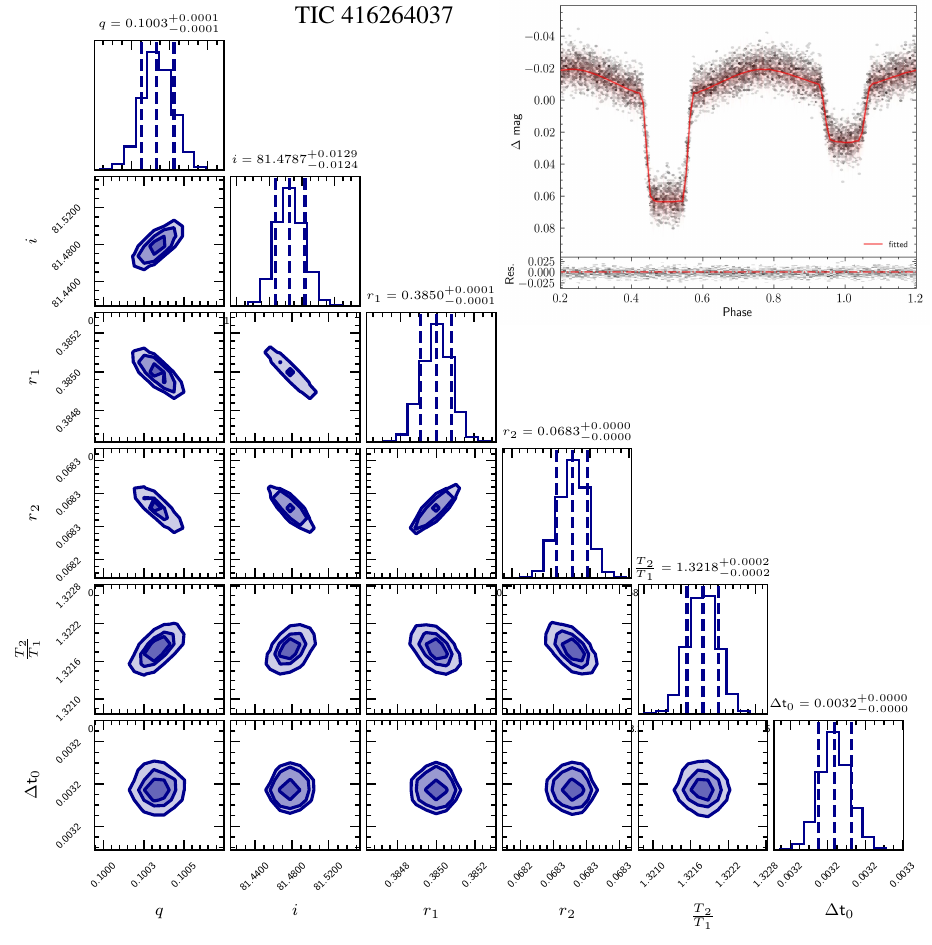}
  }
  \centering
  \caption{The light curve fitting results for TIC 149160359 (left) and TIC 416264037 (right) \citep{2020ApJ...888...49W}. The posterior distributions and the uncertainties of the fitting parameters are shown in blue histograms. In the top right corner, the grayscale plot in black shades represents the observed light curve from TESS survey, where darker shades indicate a higher concentration of data points within each grid cell. The red solid line represents the model-generated light curve derived using the best-fit parameters. Red error bars denote the scatter in the observational data, while the corresponding residuals are displayed as a grayscale plot in the lower section of the upper panel.}\label{fig:LCfitting}
\end{figure*}

According to our previous work \citep{2024ApJS..270...20X}, we build a fast light curve fitting model for EL CVn-type light curves to expedite the fitting process. The training dataset is constructed by using PHOEBE \citep{2005ApJ...628..426P,2016ApJS..227...29P,2020ApJS..247...63J}.
In this framework, the more massive A/F-type star is designated as primary star (star ``1"), while the pre-He white dwarf is labeled as secondary star (star ``2"). We adopt the following parameter ranges \citep{2018MNRAS.475.2560V, 2024NewA..10702153P} when generating mock light curves for EL CVn-type binaries with PHOEBE: (1) the mass ratio ($q$=$M_{2}$/$M_{1}$) ranges from 0.01 to 0.2; (2) the inclination is between 60$^{\circ}$ and 90$^{\circ}$; (3) the effective temperature of primary star ($T\rm_{1}$) is in the range of 6500 K $\sim$ 10000 K, which corresponding to the range of A/F type main sequence stars; (4) the effective temperature of secondary star ($T\rm_{2}$) is within the range of 7900 K $\sim$ 17000 K; (5) the relative radii of two components are generated in the range of 0.01$R\rm _{L_{1}}$ $\sim$ $R\rm _{L_{1}}$ and 0.01$R\rm _{L_{2}}$ $\sim$ $R\rm _{L_{2}}$, where $R\rm _{L_{1}}$ and $R\rm _{L_{2}}$ are the Roche lobe radii of two components. And the Roche lobe radii are constrained by mass ratio as follows \citep{1983ApJ...268..368E}:
    \begin{equation}\label{eq:RL1}
        \frac{R_{\mathrm{L_{1}}}}{a} = \frac{0.49q^{-2/3}}{0.6q^{-2/3}+\mathrm{ln}(1+q^{-1/3})}.
    \end{equation}
In Eq.\ref{eq:RL1}, $a$ is the semi-major axis, and the mass ratio $q$ is defined as $M_{2}$/$M_{1}$.
Following the law presented by \citep{1924MNRAS..84..665V,1967ZA.....65...89L,1969AcA....19..245R}, when generating the light curves for EL CVn-type binaries in PHOEBE, we adopt the values of gravity darkening coefficients \emph{g}$_{1,2}$ and reflection coefficients (Albedos, \emph{A}$_{1,2}$) with the temperatures of the components.

Similar to previous work, we also employ a Multi-Layer Perceptron (MLP) network to build the light curve fitting model for EL CVn-type binaries. Based on the trained neural network (NN) model, we utilize the Markov Chain Monte Carlo (MCMC) algorithm \citep{2013PASP..125..306F} \footnote{\url{http://dan.iel.fm/emcee}} combined with the DBSCAN clustering method to estimate the uncertainties of the fitting parameters. The initial ranges for these parameters are consistent with those used in the light curve fitting model for EL CVn-type binaries. The number of walkers and iterations for the MCMC process are set to 100 and 800, respectively. Subsequently, we apply DBSCAN clustering to the MCMC output to identify the parameter set that yields the best-fitting light curve in comparison to the observational data, which we adopt as the final solution. The goodness of fit is assessed using the the R-squared value ($\mathcal{R}^{2}$), which is defined as Eq.\ref{eq:R2}: 
\begin{equation}\label{eq:R2}
   \mathcal{R}^{2}=1-\frac{\sum^{n}_{i=1}(y_{i}-f(x_{i}))^{2}}{\sum^{n}_{i=1}(y_{i}- \overline{y})^{2}}
\end{equation}
where $f(x_{i})$ is the light curve generated by the model, $y$ represents the observed light curve. $n$ denotes the total number of data points in the light curve.

Fig.~\ref{fig:LCfitting} shows the light curve fitting results for TIC 149160359 and TIC 416264037 \citep{2020ApJ...888...49W}. In Fig.~\ref{fig:LCfitting}, the posterior distributions and the uncertainties of the fitting parameters $q$, $i$, $r_1$, $r_2$, and $T_2/T_1$ for the EL CVn systems are shown. Additionally, $\Delta T_0$ refers to the bias for the zero-point of phase. In the top right corner, a comparison between the observed and modeled light curves is presented. The grayscale plot (rendered in black shades) represents the observed light curve, while the red solid line corresponds to the model-generated light curve using the best-fit parameters. The red error bars illustrate the scatter in the observational data. It can be seen that the method provides an good fit to the observed light curve. And the parameters are also generally consistent with those provided in \citet{2020ApJ...888...49W} (see in Table.\ref{tab:table1}).

\subsection{SED fitting}

\begin{figure*}[t]
  \centering
    \subfigure{
   \includegraphics[scale=0.5]{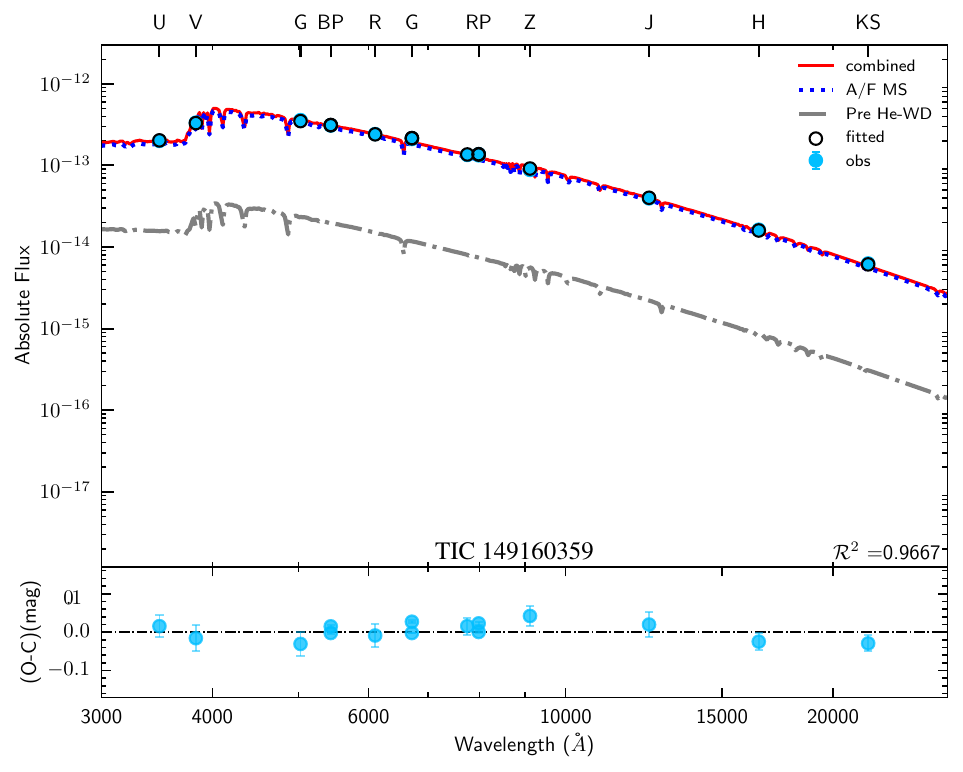}}
  \subfigure{
   \includegraphics[scale=0.5]{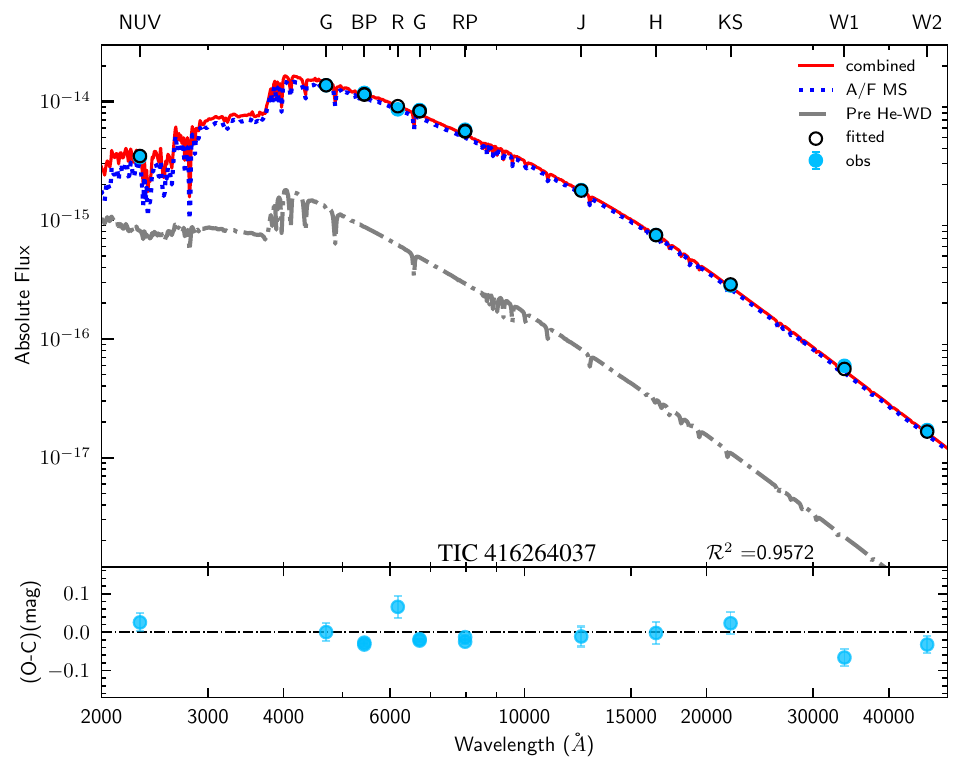}
  }
  \centering
  \caption{The SED fitting results for TIC 149160359 (left) and TIC 416264037 (right) \citep{2020ApJ...888...49W}. The blue filled circles and black open circles represent the observed and fitted photometric values, respectively. The blue dashed line is the SED of the A/F-type main sequence star, and the gray dash-dotted line shows the SED for the pre-He WD. The combined SED is shown in red solid line.}\label{fig:SEDfitting}
\end{figure*}

By modeling the light curves, we constrain the relative parameters of the two components, including the mass ratio ($q$), inclination ($i$), relative radii of the components ($r_1$, $r_2$), and the temperature ratio between the components ($T_2/T_1$). However, due to the limitations of using a single-band light curve, we cannot accurately determine their absolute effective temperatures. Gaia DR3 has provided astrometric and broad-band photometric measurements for approximately 1.5 billion objects \citep{2023A&A...674A...1G}. By integrating multi-band photometric data from GALEX, 2MASS, WISE, SkyMapper and APASS, we can perform spectral energy distribution (SED) fitting to derive more accurate estimates of the effective temperatures \citep{2011MNRAS.418.1156M, 2018MNRAS.475.2560V}. 

For the SED fitting, we first construct a cubic interpolator using the ATLAS library \citep{1979ApJS...40....1K}. During the fitting process, we utilize \textit{Speedyfit}\footnote{\url{https://speedyfit.readthedocs.io/en/stable/index.html}} package to download photometric data from the aforementioned large surveys, as well as Vizier (if available). Then any photometric data points with observational errors $>$ 0.1 ``\texttt{mag}'' are excluded. In the fitting procedure, the surface gravities (\logg) for the A/F-type main sequence stars and the pre-He WD components are set to $4.0$ and $5.0$, respectively \citep{2018MNRAS.475.2560V}. The extinction prior is derived from the 3D dust map provided by \textit{dustmaps bayestar} \citep{2019ApJ...887...93G}. For targets where 3D dust maps are unavailable due to distance or positional constraints, we utilize the two-dimensional \textit{SFD} maps \citep{1998ApJ...500..525S} to compute the extinction prior. The effective temperature ratios and radius ratios are fixed based on the results from the light curve modeling. Consequently, the parameters we fit in the SED analysis are the effective temperatures (T$_{1}$ and T$_{2}$) and extinction values of the two components.

The observed photometry represents the total flux contribution from both the A/F-type main sequence star and the pre-He white dwarf. Using the MCMC method along with the cubic interpolator, we calculate the flux contribution of each component in each band, based on varying effective temperatures, \logg\,, and extinction ($E(B-V)$). By adjusting these parameters, we aim to derive a set of effective temperatures, \logg\,, and extinction that best fit the observed SED.

Fig.~\ref{fig:SEDfitting} shows the SED fitting results for TIC 149160359 (left) and TIC 416264037 (right). In these plots, the blue circles correspond to the observed photometric values, while the black open circles indicate the fitted values. The blue dashed line represents the SED of the A/F-type main sequence star, and the gray dotted line shows the SED contribution from the pre-He white dwarf (WD). The red solid line illustrates the total SED fit. The residuals between the observed and fitted values are displayed in the lower panels. As shown, the A/F-type main sequence star contributes the majority of the photometric flux from the near-ultraviolet (NUV) to the optical bands, whereas the pre-He WD's contribution is prominent primarily in the far-ultraviolet (FUV) or shorter wavelengths. 

\begin{figure}[h]
  \centering
   \includegraphics[scale=0.7]{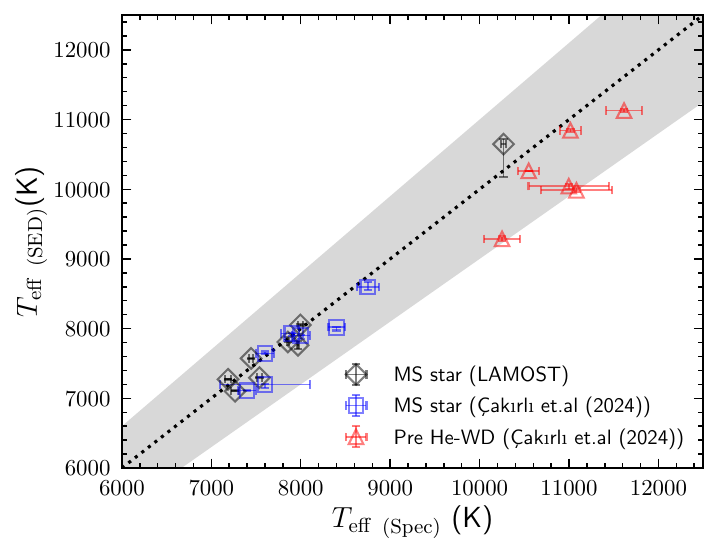}
  \centering
  \caption{Comparison of effective temperatures (\teff) derived from SED fitting and spectroscopic measurements. The $x$-axis represents effective temperatures obtained by spectra, while the $y$-axis shows results from the SED fitting. Blue open squares and triangles represent the comparisons with the effective temperatures derived from high-resolution double-lined spectroscopic measurements \citep{2024MNRAS.533.2058C}  for the A/F-type main-sequence star and the pre-He WD, respectively. Black diamonds indicate comparisons with effective temperatures derived from LAMOST survey (DR11 and DR12). The gray shaded region highlights the area with deviations of ±10\%.}\label{fig:teffcompare}
\end{figure}

To assess the reliability of the effective temperatures derived from SED fitting, we compare them with those obtained from spectra. \citet{2024MNRAS.533.2058C} have fitted the parameters of nine EL CVn systems using high-resolution spectra (resolution: 50,000, wavelength: 300 $\rm{nm}$ to 1100 $\rm{nm}$) obtained with the ESO VLT/UT2 (Kueyen\footnote{\url{https://www.eso.org/public/teles-instr/paranal-observatory/vlt/}}) telescope \citep{2000SPIE.4008..534D}. $7$ of their samples overlap with ours. Additionally, the LAMOST survey \citep{2012RAA....12.1197C,2012RAA....12..723Z,2015RAA....15.1095L,2020arXiv200507210L}, which provides both low-resolution (resolution: 1800, wavelength: 3690–9100 \AA) and medium-resolution (resolution: 7500, wavelength: 4950–5350 \AA and 6300–6800 \AA) spectra, has made available more than 10 million low-resolution spectra and over 50 million medium-resolution spectra in LAMOST DR11 and DR12. By cross-matching our selected EL CVn candidates with LAMOST DR11 and DR12 stellar parameters, we identify 8 common sources.

The comparison between the effective temperatures derived from SED fitting and those obtained from spectra is presented in Fig.\ref{fig:teffcompare}. In this figure, the $x$-axis represents the effective temperatures measured from spectra, while the $y$-axis shows the effective temperatures derived from SED fitting. The blue open squares and red triangles represent the effective temperatures determined from our SED fitting of the A/F-type main-sequence star and the pre-He white dwarf, respectively. These values are compared to the temperatures obtained by \citet{2024MNRAS.533.2058C} through double-lined high-resolution spectra and light curve modeling. The black diamonds illustrate the comparison between the effective temperatures from SED fitting and those measured from LAMOST spectra. Since the LAMOST spectra do not cover the near-ultraviolet and ultraviolet regions, the observed spectra are predominantly influenced by the A/F-type main-sequence star. The shaded region represents the range of temperature deviations of $\pm$10\%.

As shown in the Fig.~\ref{fig:teffcompare}, it illustrates that the effective temperatures of the A/F-type main-sequence stars derived from our SED fitting are aligned well with the spectroscopic measurements, with the mean difference $<$ 100 $K$. For the pre-He WD, the mean effective temperature difference is around 800 $K$ ($\sim$10\%). The effective temperature of the pre-He WD is calculated using the temperature of the A/F-type star and the temperature ratio, with the error propagation handled through Monte Carlo sampling.

\subsection{Mass and Radius}

\begin{figure}[t]
  \centering
   \includegraphics[scale=0.55]{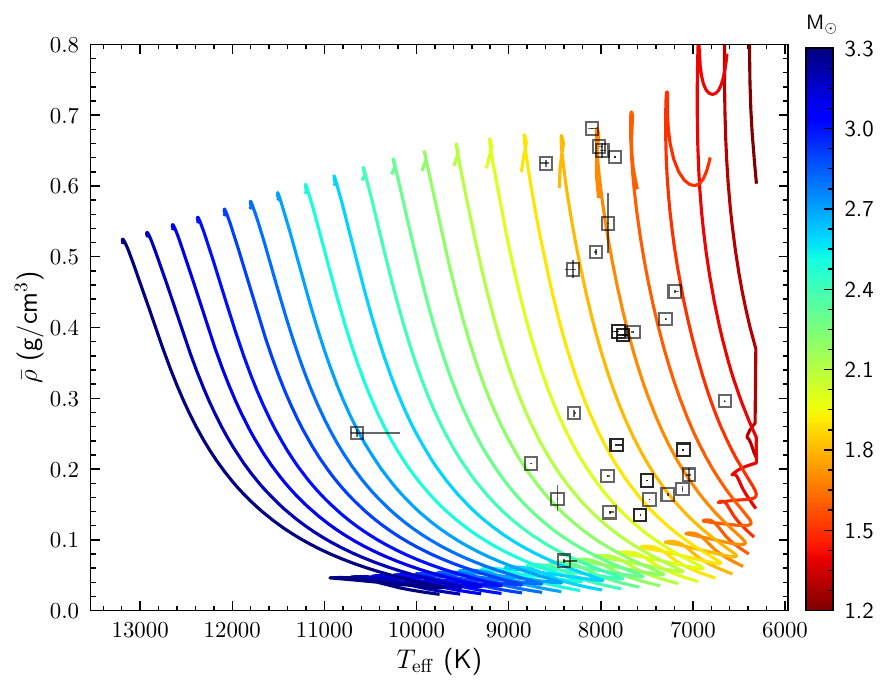}
  \centering
  \caption{The $\bar{\rho}$ - \teff\, distribution of observed EL CVn systems. The different colored lines represent the $\bar{\rho}$ - \teff\, grid for main-sequence stars derived from the MIST models ([M/H]=0, \citet{2011ApJS..192....3P,2013ApJS..208....4P,2015ApJS..220...15P,2016ApJS..222....8D,2016ApJ...823..102C}), while the open squares shown the samples from TESS survey.}\label{fig:MIST}
\end{figure}
The uncertainty in mass measurement using parallax to determine stellar luminosity can $>$ 20\% \citep{2024ApJS..270...20X}, making this method unsuitable for accurately estimating the masses of low-mass pre-He WDs. A relationship exists between the stellar mean density ($\bar{\rho}$) and effective temperature for main-sequence stars. Similar to the methods used by \citet{2013ApJ...777...77Z, 2015ApJ...803...82R} and \citet{2020ApJ...888...49W}, we apply this relationship to estimate the mass of the A/F-type main-sequence star. The stellar mean density can be derived from the relative radius ($r_{1}$), mass ratio ($q$), and orbital period ($P$), as expressed in Eq.\ref{Equ:rho}, which is derived by combining \textit{Kepler's} third law with the formula for calculating density.

\begin{equation}\label{Equ:rho}
\begin{aligned}
    \bar{\rho_{1}}=\frac{3\pi}{GP^2(r_{1}^{3}(1+q))}
    \end{aligned}
\end{equation}

Fig.\ref{fig:MIST} illustrates the distribution of $\bar{\rho}$ - \teff\, for the EL CVn systems detected in the TESS survey. The $\bar{\rho}$ - \teff\, grid for main-sequence stars is derived from the MIST models \citep{2011ApJS..192....3P,2013ApJS..208....4P,2015ApJS..220...15P,2016ApJS..222....8D,2016ApJ...823..102C}, with a mass grid interval of 0.01M$\odot$. The solid colored lines represent the $\bar{\rho}$ - \teff\, relationship for different masses, while the black open rectangles mark the EL CVn systems identified in TESS. Based on the position of each sample in the $\bar{\rho}$ - \teff\, diagram, we determine the mass of the A/F-type main-sequence star in these systems. The uncertainty in mass is estimated using Monte Carlo sampling. By combining the mass ratio obtained from the light curve fitting, we further derive the mass of the pre-He WD. The radius is then calculated using $\bar{\rho_{2}}=M_{2}/(4\pi R_{2}^{3}/3)$. Moreover, due to the absence of observed metallicity measurements, we use the $\bar{\rho}$ - \teff\, relationship corresponding to solar metallicity ([M/H]=0) only.

\begin{figure}[h]
  \centering
   \includegraphics[scale=0.7]{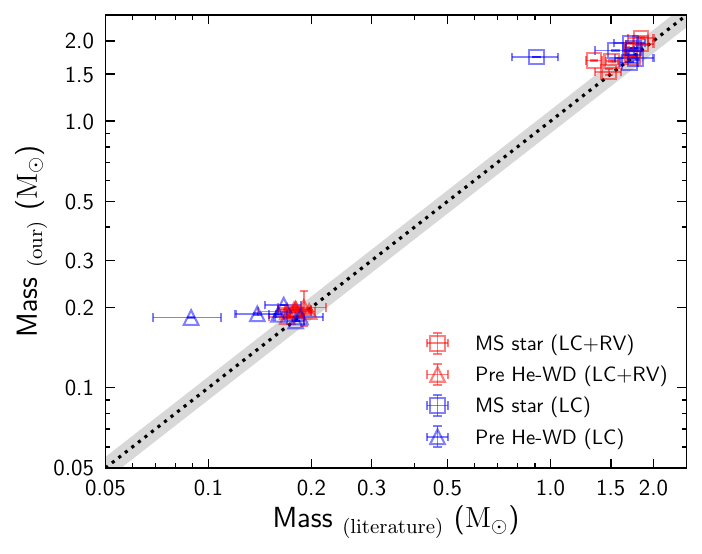}
  \centering
  \caption{The comparison of mass measurements. The $x$-axis shows the previously known values and the $y$-axis represents our measured results. The red squares and triangles indicate results obtained from both double-lined spectroscopy and light curve fitting, while the blue squares and triangles represent results derived solely from light curve fitting. The gray shaded region highlights the area with deviations of ±10\%.}\label{fig:masscompare}
\end{figure}

Similarly, we compare the masses obtained from other literature (see Sec.~\ref{sec_intro}) with those derived from our analysis. Fig.~\ref{fig:masscompare} shows the mass comparison, where the red open rectangles and triangles represent masses measured using both spectroscopy and light curve fitting, while the blue rectangles and triangles indicate masses derived from light curve fitting alone. The $x$-axis represents the masses collected from the literature, while the $y$-axis shows the masses measured in our analysis. It can be seen that the mass measurements obtained using our method are generally consistent with those reported in the literature and align well with the masses derived from spectroscopic and light curve analyses. The target with the larger discrepancy is due to that their effective temperature was fitted using only the light curve, which could result in a deviation between the absolute temperature values measured through SED fitting and those obtained in their analysis. The gray shaded region represents the range of temperature deviations of $\pm$10\%. Moreover, the complete parameters of 29 EL CVn systems identified from TESS survey are listed in Appendix.~\ref{AppendixA}.

\section{Properties} \label{sec:discuss}
\begin{figure}[t]
  \centering
   \includegraphics[scale=0.55]{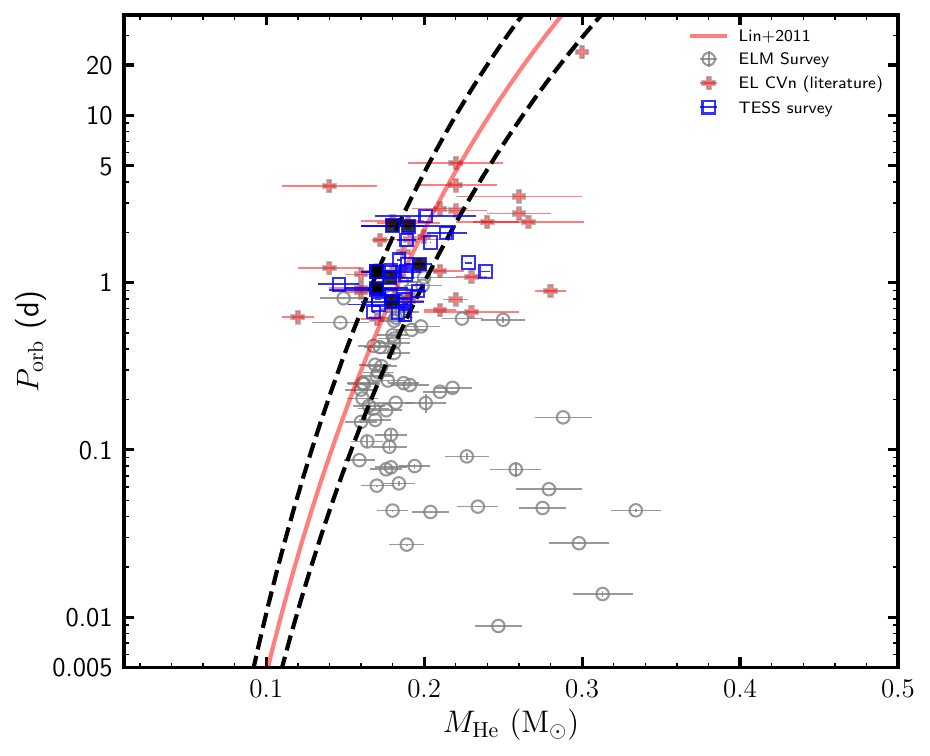}
  \centering
  \caption{The period-mass distribution. The red solid line representing the period–mass relation proposed by \citet{2011ApJ...732...70L}, and the black dashed lines marking the ±10\% uncertainty range around this relation. Gray open circles depict the extremely low-mass (ELM) white dwarf samples compiled by \citet{2019ApJ...871..148L}. The red ``plus" symbols denote the previously known EL CVn systems (refer to Sec.~\ref{sec_intro}), while the blue open squares indicate EL CVn candidates identified from the TESS survey.}\label{fig:PMhe}
\end{figure}

The period-mass distribution plays a crucial role in understanding the evolutionary channels of binary star systems. Mass-transfer models indicate a relationship between the mass of the WD and its final orbital period. To investigate this relationship, we analyze the mass-period distribution of EL CVn systems. Fig. \ref{fig:PMhe} presents the period-mass distribution for EL CVn systems. In Fig. \ref{fig:PMhe}, the gray open circles represent the ELM white dwarf samples compiled by \citet{2019ApJ...871..148L}, while the red ``plus" corresponds to the EL CVn samples from other literature (see Sec.~\ref{sec_intro}). The blue open squares are the EL CVn candidates obtained from TESS survey. The red solid line is the period–mass relation proposed by \citet{2011ApJ...732...70L}, and the black dashed lines indicate the ±10\% uncertainties associated with this relation. Considering the observational uncertainties, it is evident that the new and previously known EL CVn systems are located near the red solid line and the two dashed lines, this alignment suggests that the EL CVn-systems evolve through the Roche lobe overflow (RL) channel and undergo stable mass transfer. 

\begin{figure}[h]
    \subfigure{
   \includegraphics[scale=0.49]{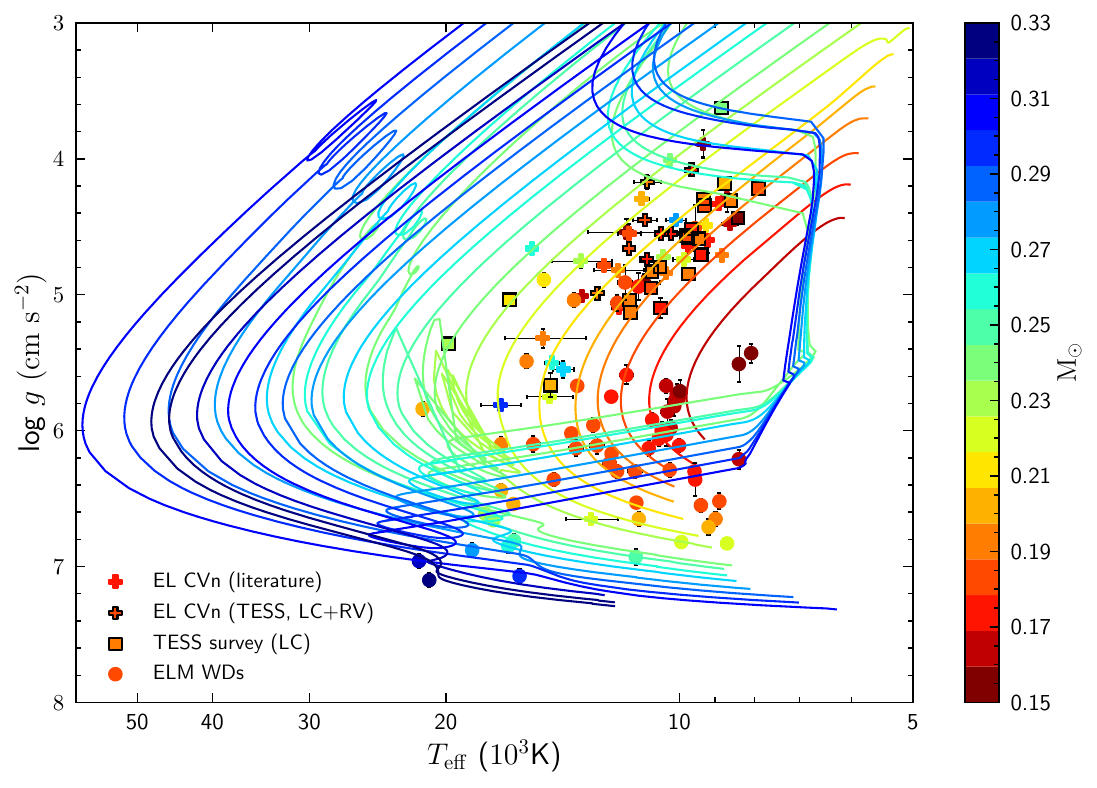}}
  \caption{\teff\,-\logg\, diagram of EL CVn systems. The colored solid lines depict the evolutionary tracks of Helium White Dwarfs (He WDs) presented by \citet{2019ApJ...871..148L}. The filled circles represent extremely low-mass white dwarfs compiled by \citet{2019ApJ...871..148L}. The filled squares outlined in black correspond to EL CVn binaries identified in the TESS survey. Additionally, filled plus symbols indicate EL CVn systems sourced from the literature, with those having black edges representing systems characterized by double-lined spectra and high-precision light curves.}\label{fig:tefflogg}
\end{figure}

Fig.\ref{fig:tefflogg} shows the \teff\,-\logg\, diagram of the systems. In Fig.\ref{fig:tefflogg}, the colored solid lines represent the evolutionary tracks of Helium White Dwarfs (He WDs) with masses ranging from 0.145 to 0.335 $M_{\odot}$ \citep{2019ApJ...871..148L}. The filled circles denote extremely low-mass (ELM) white dwarfs compiled by \citet{2019ApJ...871..148L}. The filled squares with black edges correspond to the EL CVn binaries identified from the TESS survey. The filled "plus" symbols indicate EL CVn systems gathered from the literature, with those marked by black edges representing systems characterized by double-lined spectra and high-precision light curves. As shown in Fig. \ref{fig:tefflogg}, the observational EL CVn systems are consistent with the evolutionary tracks of the models, further supporting the scenario of stable mass transfer. For the few outliers identified, the metallicity information would allow us to better constrain their parameters.

Furthermore, by comparing the theoretical evolutionary tracks with the observed data, it is evident that the He WDs in EL CVn systems are in an earlier evolutionary stage than typical ELM  WDs. The white dwarfs in EL CVn systems have recently completed mass transfer and are still undergoing contraction, as indicated by their relatively higher surface gravity and lower effective temperature compared to standard ELM WDs. Over time, the pre-He WDs in EL CVn systems will cool, leading to an increase in both surface gravity and effective temperature until they align with the typical parameter range of ELM WDs. These EL CVn systems further support the stability of the mass transfer process, enhancing our understanding of the stable mass transfer channels that contribute to the formation of such white dwarfs.

\section{summary} \label{sec:summary}
EL CVn-type binaries are crucial for understanding the process of mass transfer in binary systems and the formation of extremely low-mass white dwarfs. In this study, we identified 29 EL CVn-type binary systems from the TESS eclipsing binary catalog (sectors 1-65). These systems consist of smaller, hotter pre-He WDs and more luminous A/F main-sequence stars, and they were selected based on their characteristic, deeper, ``boxy'' shaped secondary eclipses. Next, using light curves simulated by PHOEBE for EL CVn-type binaries, we trained a light curve fitting model with an MLP model to fit their light curves. This allowed us to determine the mass ratio ($q$), inclination ($i$), relative radii ($r_{1,2}$), and temperature ratio ($T_{2}/T_{1}$). With the effective temperature and radius ratios fixed, we then used the parallax provided by Gaia to calculate extinction and used it as a prior for fitting the SED to solve for the absolute effective temperature values of the two components. Based on the \teff\,-$\rho$ relation of main-sequence stars, we derived their complete physical parameters, including mass and radius. The temperatures and masses obtained using this method align well with those derived from spectral data and dynamical mass measurements. Subsequently, we analyzed these systems in terms of mass-period and ELM WD evolutionary tracks. The results show that the observational data match the theoretical curves well and indicate that these systems have formed through stable mass transfer channels, consistent with \citet{2017MNRAS.467.1874C}.  

Additionally, in this study, we restrict our identification of EL CVn systems to those based on the 2-minute and 30-minute light curves obtained from the TESS survey. This catalog represents a subset of the total EL CVn systems in the TESS survey. Our current sample comprises approximately one-third of the EL CVn systems with complete parameters. Furthermore, the \teff\,-\logg\, diagram indicates that two additional cases have been identified, which exhibit relatively higher \logg\, pre-He WDs. The identified EL CVn systems are significant for the study of binary mass transfer and the formation and evolution of white dwarfs. Recently, pulsations of EL CVn systems have also become a hot topic, and the discovery of these samples provides substantial observational support for the study of pulsations in EL CVn-type binaries. Furthermore, our method can be applied to identify additional EL CVn systems in the TESS survey by combining the EB catalogs from other ground-based surveys (e.g., ZTF, ASAS-SN) with the TESS light curves.

\section{acknowledgement}
This work is supported by the
National Natural Science Foundation of China (NSFC) with grant Nos.12288102, 12125303, 12090040/3, the National Key R\&D Program of China (grant No.2021YFA1600401/ 2021YFA1600403), the NFSC (grant Nos.12303106, 12473034, 12103086, 12373037), the Postdoctoral Fellowship Program of CPSF (No.GZC20232976), Yunnan Fundamental Research Projects (grant Nos.202401AT070139, 202101AU070276), the International Centre of Supernovae, Yunnan Key Laboratory (No.202302AN360001) and the Yunnan Revitalization Talent Support Program—Science \& Technology Champion Project (No.202305AB350003). This work is also supported by the China Manned Space Project of No.CMS-CSST-2021-A10. We acknowledge the TESS mission provided by NASA's Science Mission Directorate. We also acknowledge the use of public TESS data from pipelines at the TESS Science Office and at the TESS Science Processing Operations Center. We thank the Gaia Data Processing and Analysis Consortium (DPAC) for their substantial contributions in producing and releasing high-quality data.

\bibliography{sample631}{}

\begin{thebibliography}{}
\expandafter\ifx\csname natexlab\endcsname\relax\def\natexlab#1{#1}\fi
\providecommand{\url}[1]{\href{#1}{#1}}
\providecommand{\dodoi}[1]{doi:~\href{http://doi.org/#1}{\nolinkurl{#1}}}
\providecommand{\doeprint}[1]{\href{http://ascl.net/#1}{\nolinkurl{http://ascl.net/#1}}}
\providecommand{\doarXiv}[1]{\href{https://arxiv.org/abs/#1}{\nolinkurl{https://arxiv.org/abs/#1}}}

\bibitem[{{Albareti} {et~al.}(2017){Albareti}, {Allende Prieto}, {Almeida}, {Anders}, {Anderson}, {Andrews}, {Arag{\'o}n-Salamanca}, {Argudo-Fern{\'a}ndez}, {Armengaud}, {Aubourg}, {Avila-Reese}, {Badenes}, {Bailey}, {Barbuy}, {Barger}, {Barrera-Ballesteros}, {Bartosz}, {Basu}, {Bates}, {Battaglia}, {Baumgarten}, {Baur}, {Bautista}, {Beers}, {Belfiore}, {Bershady}, {Bertran de Lis}, {Bird}, {Bizyaev}, {Blanc}, {Blanton}, {Blomqvist}, {Bolton}, {Borissova}, {Bovy}, {Brandt}, {Brinkmann}, {Brownstein}, {Bundy}, {Burtin}, {Busca}, {Camacho Chavez}, {Cano D{\'\i}az}, {Cappellari}, {Carrera}, {Chen}, {Cherinka}, {Cheung}, {Chiappini}, {Chojnowski}, {Chuang}, {Chung}, {Cirolini}, {Clerc}, {Cohen}, {Comerford}, {Comparat}, {Correa do Nascimento}, {Cousinou}, {Covey}, {Crane}, {Croft}, {Cunha}, {Darling}, {Davidson}, {Dawson}, {Da Costa}, {Da Silva Ilha}, {Deconto Machado}, {Delubac}, {De Lee}, {De la Macorra}, {De la Torre}, {Diamond-Stanic}, {Donor}, {Downes}, {Drory}, {Du}, {Du Mas des Bourboux}, {Dwelly},
  {Ebelke}, {Eigenbrot}, {Eisenstein}, {Elsworth}, {Emsellem}, {Eracleous}, {Escoffier}, {Evans}, {Falc{\'o}n-Barroso}, {Fan}, {Favole}, {Fernandez-Alvar}, {Fernandez-Trincado}, {Feuillet}, {Fleming}, {Font-Ribera}, {Freischlad}, {Frinchaboy}, {Fu}, {Gao}, {Garcia}, {Garcia-Dias}, {Garcia-Hern{\'a}ndez}, {Garcia P{\'e}rez}, {Gaulme}, {Ge}, {Geisler}, {Gillespie}, {Gil Marin}, {Girardi}, {Goddard}, {Gomez Maqueo Chew}, {Gonzalez-Perez}, {Grabowski}, {Green}, {Grier}, {Grier}, {Guo}, {Guy}, {Hagen}, {Hall}, {Harding}, {Harley}, {Hasselquist}, {Hawley}, {Hayes}, {Hearty}, {Hekker}, {Hernandez Toledo}, {Ho}, {Hogg}, {Holley-Bockelmann}, {Holtzman}, {Holzer}, {Hu}, {Huber}, {Hutchinson}, {Hwang}, {Ibarra-Medel}, {Ivans}, {Ivory}, {Jaehnig}, {Jensen}, {Johnson}, {Jones}, {Jullo}, {Kallinger}, {Kinemuchi}, {Kirkby}, {Klaene}, {Kneib}, {Kollmeier}, {Lacerna}, {Lane}, {Lang}, {Laurent}, {Law}, {Leauthaud}, {Le Goff}, {Li}, {Li}, {Li}, {Li}, {Liang}, {Liang}, {Lima}, {Lin}, {Lin}, {Lin}, {Liu}, {Long}, {Lucatello},
  {MacDonald}, {MacLeod}, {Mackereth}, {Mahadevan}, {Maia}, {Maiolino}, {Majewski}, {Malanushenko}, {Malanushenko}, {Mallmann}, {Manchado}, {Maraston}, {Marques-Chaves}, {Martinez Valpuesta}, {Masters}, {Mathur}, {McGreer}, {Merloni}, {Merrifield}, {M{\'e}sz{\'a}ros}, {Meza}, {Miglio}, {Minchev}, {Molaverdikhani}, {Montero-Dorta}, {Mosser}, {Muna}, {Myers}, {Nair}, {Nandra}, {Ness}, {Newman}, {Nichol}, {Nidever}, {Nitschelm}, {O'Connell}, {Oravetz}, {Oravetz}, {Pace}, {Padilla}, {Palanque-Delabrouille}, {Pan}, {Parejko}, {Paris}, {Park}, {Peacock}, {Peirani}, {Pellejero-Ibanez}, {Penny}, {Percival}, {Percival}, {Perez-Fournon}, {Petitjean}, {Pieri}, {Pinsonneault}, {Pisani}, {Prada}, {Prakash}, {Price-Jones}, {Raddick}, {Rahman}, {Raichoor}, {Barboza Rembold}, {Reyna}, {Rich}, {Richstein}, {Ridl}, {Riffel}, {Riffel}, {Rix}, {Robin}, {Rockosi}, {Rodr{\'\i}guez-Torres}, {Rodrigues}, {Roe}, {Roman Lopes}, {Rom{\'a}n-Z{\'u}{\~n}iga}, {Ross}, {Rossi}, {Ruan}, {Ruggeri}, {Runnoe}, {Salazar-Albornoz}, {Salvato},
  {Sanchez}, {Sanchez}, {Sanchez-Gallego}, {Santiago}, {Schiavon}, {Schimoia}, {Schlafly}, {Schlegel}, {Schneider}, {Sch{\"o}nrich}, {Schultheis}, {Schwope}, {Seo}, {Serenelli}, {Sesar}, {Shao}, {Shetrone}, {Shull}, {Silva Aguirre}, {Skrutskie}, {Slosar}, {Smith}, {Smith}, {Sobeck}, {Somers}, {Souto}, {Stark}, {Stassun}, {Steinmetz}, {Stello}, {Storchi Bergmann}, {Strauss}, {Streblyanska}, {Stringfellow}, {Suarez}, {Sun}, {Taghizadeh-Popp}, {Tang}, {Tao}, {Tayar}, {Tembe}, {Thomas}, {Tinker}, {Tojeiro}, {Tremonti}, {Troup}, {Trump}, {Unda-Sanzana}, {Valenzuela}, {Van den Bosch}, {Vargas-Maga{\~n}a}, {Vazquez}, {Villanova}, {Vivek}, {Vogt}, {Wake}, {Walterbos}, {Wang}, {Wang}, {Weaver}, {Weijmans}, {Weinberg}, {Westfall}, {Whelan}, {Wilcots}, {Wild}, {Williams}, {Wilson}, {Wood-Vasey}, {Wylezalek}, {Xiao}, {Yan}, {Yang}, {Ybarra}, {Yeche}, {Yuan}, {Zakamska}, {Zamora}, {Zasowski}, {Zhang}, {Zhao}, {Zhao}, {Zheng}, {Zheng}, {Zhou}, {Zhu}, {Zinn}, \& {Zou}}]{2017ApJS..233...25A}
{Albareti}, F.~D., {Allende Prieto}, C., {Almeida}, A., {et~al.} 2017, \apjs, 233, 25, \dodoi{10.3847/1538-4365/aa8992}

\bibitem[{{Bianchi} {et~al.}(2014){Bianchi}, {Conti}, \& {Shiao}}]{2014AdSpR..53..900B}
{Bianchi}, L., {Conti}, A., \& {Shiao}, B. 2014, Advances in Space Research, 53, 900, \dodoi{10.1016/j.asr.2013.07.045}

\bibitem[{{Breton} {et~al.}(2012){Breton}, {Rappaport}, {van Kerkwijk}, \& {Carter}}]{2012ApJ...748..115B}
{Breton}, R.~P., {Rappaport}, S.~A., {van Kerkwijk}, M.~H., \& {Carter}, J.~A. 2012, \apj, 748, 115, \dodoi{10.1088/0004-637X/748/2/115}

\bibitem[{{Caldwell} {et~al.}(2020){Caldwell}, {Tenenbaum}, {Twicken}, {Jenkins}, {Ting}, {Smith}, {Hedges}, {Fausnaugh}, {Rose}, \& {Burke}}]{2020RNAAS...4..201C}
{Caldwell}, D.~A., {Tenenbaum}, P., {Twicken}, J.~D., {et~al.} 2020, Research Notes of the American Astronomical Society, 4, 201, \dodoi{10.3847/2515-5172/abc9b3}

\bibitem[{{Carter} {et~al.}(2011){Carter}, {Rappaport}, \& {Fabrycky}}]{2011ApJ...728..139C}
{Carter}, J.~A., {Rappaport}, S., \& {Fabrycky}, D. 2011, \apj, 728, 139, \dodoi{10.1088/0004-637X/728/2/139}

\bibitem[{{{\c{C}}ak{\i}rl{\i}} {et~al.}(2024){{\c{C}}ak{\i}rl{\i}}, {Hoyman}, \& {{\"O}zdarcan}}]{2024MNRAS.533.2058C}
{{\c{C}}ak{\i}rl{\i}}, {\"O}., {Hoyman}, B., \& {{\"O}zdarcan}, O. 2024, \mnras, 533, 2058, \dodoi{10.1093/mnras/stae1948}

\bibitem[{{Chen} {et~al.}(2017){Chen}, {Maxted}, {Li}, \& {Han}}]{2017MNRAS.467.1874C}
{Chen}, X., {Maxted}, P.~F.~L., {Li}, J., \& {Han}, Z. 2017, \mnras, 467, 1874, \dodoi{10.1093/mnras/stx115}

\bibitem[{{Chen} {et~al.}(2022){Chen}, {Ding}, {Cheng}, {Zhang}, {Li}, {Ji}, {Xiong}, {Li}, \& {Luo}}]{2022ApJS..263...34C}
{Chen}, X., {Ding}, X., {Cheng}, L., {et~al.} 2022, \apjs, 263, 34, \dodoi{10.3847/1538-4365/aca284}

\bibitem[{{Choi} {et~al.}(2016){Choi}, {Dotter}, {Conroy}, {Cantiello}, {Paxton}, \& {Johnson}}]{2016ApJ...823..102C}
{Choi}, J., {Dotter}, A., {Conroy}, C., {et~al.} 2016, \apj, 823, 102, \dodoi{10.3847/0004-637X/823/2/102}

\bibitem[{{Cui} {et~al.}(2012){Cui}, {Zhao}, {Chu}, {Li}, {Li}, {Zhang}, {Su}, {Yao}, {Wang}, {Xing}, {Li}, {Zhu}, {Wang}, {Gu}, {Luo}, {Xu}, {Zhang}, {Liu}, {Zhang}, {Yang}, {Cao}, {Chen}, {Chen}, {Chen}, {Chen}, {Chu}, {Feng}, {Gong}, {Hou}, {Hu}, {Hu}, {Hu}, {Jia}, {Jiang}, {Jiang}, {Jiang}, {Jin}, {Li}, {Li}, {Li}, {Liu}, {Liu}, {Lu}, {Mao}, {Men}, {Qi}, {Qi}, {Shi}, {Tang}, {Tao}, {Wang}, {Wang}, {Wang}, {Wang}, {Wang}, {Wang}, {Wang}, {Wang}, {Wang}, {Wang}, {Wang}, {Wang}, {Xu}, {Xu}, {Yang}, {Yu}, {Yuan}, {Yuan}, {Zhai}, {Zhang}, {Zhang}, {Zhang}, {Zhao}, {Zhou}, {Zhou}, {Zhu}, \& {Zou}}]{2012RAA....12.1197C}
{Cui}, X.-Q., {Zhao}, Y.-H., {Chu}, Y.-Q., {et~al.} 2012, Research in Astronomy and Astrophysics, 12, 1197, \dodoi{10.1088/1674-4527/12/9/003}

\bibitem[{{Dekker} {et~al.}(2000){Dekker}, {D'Odorico}, {Kaufer}, {Delabre}, \& {Kotzlowski}}]{2000SPIE.4008..534D}
{Dekker}, H., {D'Odorico}, S., {Kaufer}, A., {Delabre}, B., \& {Kotzlowski}, H. 2000, in Society of Photo-Optical Instrumentation Engineers (SPIE) Conference Series, Vol. 4008, Optical and IR Telescope Instrumentation and Detectors, ed. M.~{Iye} \& A.~F. {Moorwood}, 534--545, \dodoi{10.1117/12.395512}

\bibitem[{{Dotter}(2016)}]{2016ApJS..222....8D}
{Dotter}, A. 2016, \apjs, 222, 8, \dodoi{10.3847/0067-0049/222/1/8}

\bibitem[{{Eggleton}(1983)}]{1983ApJ...268..368E}
{Eggleton}, P.~P. 1983, \apj, 268, 368, \dodoi{10.1086/160960}

\bibitem[{{Faigler} {et~al.}(2015){Faigler}, {Kull}, {Mazeh}, {Kiefer}, {Latham}, \& {Bloemen}}]{2015ApJ...815...26F}
{Faigler}, S., {Kull}, I., {Mazeh}, T., {et~al.} 2015, \apj, 815, 26, \dodoi{10.1088/0004-637X/815/1/26}

\bibitem[{{Foreman-Mackey} {et~al.}(2013){Foreman-Mackey}, {Hogg}, {Lang}, \& {Goodman}}]{2013PASP..125..306F}
{Foreman-Mackey}, D., {Hogg}, D.~W., {Lang}, D., \& {Goodman}, J. 2013, \pasp, 125, 306, \dodoi{10.1086/670067}

\bibitem[{{Gaia Collaboration} {et~al.}(2023{\natexlab{a}}){Gaia Collaboration}, {Arenou}, {Babusiaux}, {Barstow}, {Faigler}, {Jorissen}, {Kervella}, {Mazeh}, {Mowlavi}, {Panuzzo}, {Sahlmann}, {Shahaf}, {Sozzetti}, {Bauchet}, {Damerdji}, {Gavras}, {Giacobbe}, {Gosset}, {Halbwachs}, {Holl}, {Lattanzi}, {Leclerc}, {Morel}, {Pourbaix}, {Re Fiorentin}, {Sadowski}, {S{\'e}gransan}, {Siopis}, {Teyssier}, {Zwitter}, {Planquart}, {Brown}, {Vallenari}, {Prusti}, {de Bruijne}, {Biermann}, {Creevey}, {Ducourant}, {Evans}, {Eyer}, {Guerra}, {Hutton}, {Jordi}, {Klioner}, {Lammers}, {Lindegren}, {Luri}, {Mignard}, {Panem}, {Randich}, {Sartoretti}, {Soubiran}, {Tanga}, {Walton}, {Bailer-Jones}, {Bastian}, {Drimmel}, {Jansen}, {Katz}, {van Leeuwen}, {Bakker}, {Cacciari}, {Casta{\~n}eda}, {De Angeli}, {Fabricius}, {Fouesneau}, {Fr{\'e}mat}, {Galluccio}, {Guerrier}, {Heiter}, {Masana}, {Messineo}, {Nicolas}, {Nienartowicz}, {Pailler}, {Riclet}, {Roux}, {Seabroke}, {Sordo}, {Th{\'e}venin}, {Gracia-Abril}, {Portell}, {Altmann},
  {Andrae}, {Audard}, {Bellas-Velidis}, {Benson}, {Berthier}, {Blomme}, {Burgess}, {Busonero}, {Busso}, {C{\'a}novas}, {Carry}, {Cellino}, {Cheek}, {Clementini}, {Davidson}, {de Teodoro}, {Nu{\~n}ez Campos}, {Delchambre}, {Dell'Oro}, {Esquej}, {Fern{\'a}ndez-Hern{\'a}ndez}, {Fraile}, {Garabato}, {Garc{\'\i}a-Lario}, {Haigron}, {Hambly}, {Harrison}, {Hern{\'a}ndez}, {Hestroffer}, {Hodgkin}, {Jan{\ss}en}, {Jevardat de Fombelle}, {Jordan}, {Krone-Martins}, {Lanzafame}, {L{\"o}ffler}, {Marchal}, {Marrese}, {Moitinho}, {Muinonen}, {Osborne}, {Pancino}, {Pauwels}, {Recio-Blanco}, {Reyl{\'e}}, {Riello}, {Rimoldini}, {Roegiers}, {Rybizki}, {Sarro}, {Smith}, {Utrilla}, {van Leeuwen}, {Abbas}, {{\'A}brah{\'a}m}, {Abreu Aramburu}, {Aerts}, {Aguado}, {Ajaj}, {Aldea-Montero}, {Altavilla}, {{\'A}lvarez}, {Alves}, {Anders}, {Anderson}, {Anglada Varela}, {Antoja}, {Baines}, {Baker}, {Balaguer-N{\'u}{\~n}ez}, {Balbinot}, {Balog}, {Barache}, {Barbato}, {Barros}, {Bartolom{\'e}}, {Bassilana}, {Becciani}, {Bellazzini},
  {Berihuete}, {Bernet}, {Bertone}, {Bianchi}, {Binnenfeld}, {Blanco-Cuaresma}, {Blazere}, {Boch}, {Bombrun}, {Bossini}, {Bouquillon}, {Bragaglia}, {Bramante}, {Breedt}, {Bressan}, {Brouillet}, {Brugaletta}, {Bucciarelli}, {Burlacu}, {Butkevich}, {Buzzi}, {Caffau}, {Cancelliere}, {Cantat-Gaudin}, {Carballo}, {Carlucci}, {Carnerero}, {Carrasco}, {Casamiquela}, {Castellani}, {Castro-Ginard}, {Chaoul}, {Charlot}, {Chemin}, {Chiaramida}, {Chiavassa}, {Chornay}, {Comoretto}, {Contursi}, {Cooper}, {Cornez}, {Cowell}, {Crifo}, {Cropper}, {Crosta}, {Crowley}, {Dafonte}, {Dapergolas}, {David}, {de Laverny}, {De Luise}, {De March}, {De Ridder}, {de Souza}, {de Torres}, {del Peloso}, {del Pozo}, {Delbo}, {Delgado}, {Delisle}, {Demouchy}, {Dharmawardena}, {Diakite}, {Diener}, {Distefano}, {Dolding}, {Enke}, {Fabre}, {Fabrizio}, {Fedorets}, {Fernique}, {Figueras}, {Fournier}, {Fouron}, {Fragkoudi}, {Gai}, {Garcia-Gutierrez}, {Garcia-Reinaldos}, {Garc{\'\i}a-Torres}, {Garofalo}, {Gavel}, {Gerlach}, {Geyer}, {Gilmore},
  {Girona}, {Giuffrida}, {Gomel}, {Gomez}, {Gonz{\'a}lez-N{\'u}{\~n}ez}, {Gonz{\'a}lez-Santamar{\'\i}a}, {Gonz{\'a}lez-Vidal}, {Granvik}, {Guillout}, {Guiraud}, {Guti{\'e}rrez-S{\'a}nchez}, {Guy}, {Hatzidimitriou}, {Hauser}, {Haywood}, {Helmer}, {Helmi}, {Sarmiento}, {Hidalgo}, {Hilger}, {H{\l}adczuk}, {Hobbs}, {Holland}, {Huckle}, {Jardine}, {Jasniewicz}, {Jean-Antoine Piccolo}, {Jim{\'e}nez-Arranz}, {Juaristi Campillo}, {Julbe}, {Karbevska}, {Khanna}, {Kordopatis}, {Korn}, {K{\'o}sp{\'a}l}, {Kostrzewa-Rutkowska}, {Kruszy{\'n}ska}, {Kun}, {Laizeau}, {Lambert}, {Lanza}, {Lasne}, {Le Campion}, {Lebreton}, {Lebzelter}, {Leccia}, {Lecoeur-Taibi}, {Liao}, {Licata}, {Lindstr{\o}m}, {Lister}, {Livanou}, {Lobel}, {Lorca}, {Loup}, {Madrero Pardo}, {Magdaleno Romeo}, {Managau}, {Mann}, {Manteiga}, {Marchant}, {Marconi}, {Marcos}, {Marcos Santos}, {Mar{\'\i}n Pina}, {Marinoni}, {Marocco}, {Marshall}, {Martin Polo}, {Mart{\'\i}n-Fleitas}, {Marton}, {Mary}, {Masip}, {Massari}, {Mastrobuono-Battisti}, {McMillan},
  {Messina}, {Michalik}, {Millar}, {Mints}, {Molina}, {Molinaro}, {Moln{\'a}r}, {Monari}, {Mongui{\'o}}, {Montegriffo}, {Montero}, {Mor}, {Mora}, {Morbidelli}, {Morris}, {Muraveva}, {Murphy}, {Musella}, {Nagy}, {Noval}, {Oca{\~n}a}, {Ogden}, {Ordenovic}, {Osinde}, {Pagani}, {Pagano}, {Palaversa}, {Palicio}, {Pallas-Quintela}, {Panahi}, {Payne-Wardenaar}, {Pe{\~n}alosa Esteller}, {Penttil{\"a}}, {Pichon}, {Piersimoni}, {Pineau}, {Plachy}, {Plum}, {Poggio}, {Pr{\v{s}}a}, {Pulone}, {Racero}, {Ragaini}, {Rainer}, {Raiteri}, {Ramos}, {Ramos-Lerate}, {Regibo}, {Richards}, {Rios Diaz}, {Ripepi}, {Riva}, {Rix}, {Rixon}, {Robichon}, {Robin}, {Robin}, {Roelens}, {Rogues}, {Rohrbasser}, {Romero-G{\'o}mez}, {Rowell}, {Royer}, {Ruz Mieres}, {Rybicki}, {S{\'a}ez N{\'u}{\~n}ez}, {Sagrist{\`a} Sell{\'e}s}, {Salguero}, {Samaras}, {Sanchez Gimenez}, {Sanna}, {Santove{\~n}a}, {Sarasso}, {Schultheis}, {Sciacca}, {Segol}, {Segovia}, {Semeux}, {Siddiqui}, {Siebert}, {Siltala}, {Silvelo}, {Slezak}, {Slezak}, {Smart}, {Snaith},
  {Solano}, {Solitro}, {Souami}, {Souchay}, {Spagna}, {Spina}, {Spoto}, {Steele}, {Steidelm{\"u}ller}, {Stephenson}, {S{\"u}veges}, {Surdej}, {Szabados}, {Szegedi-Elek}, {Taris}, {Taylor}, {Teixeira}, {Tolomei}, {Tonello}, {Torra}, {Torra}, {Torralba Elipe}, {Trabucchi}, {Tsounis}, {Turon}, {Ulla}, {Unger}, {Vaillant}, {van Dillen}, {van Reeven}, {Vanel}, {Vecchiato}, {Viala}, {Vicente}, {Voutsinas}, {Weiler}, {Wevers}, {Wyrzykowski}, {Yoldas}, {Yvard}, {Zhao}, {Zorec}, \& {Zucker}}]{2023A&A...674A..34G}
{Gaia Collaboration}, {Arenou}, F., {Babusiaux}, C., {et~al.} 2023{\natexlab{a}}, \aap, 674, A34, \dodoi{10.1051/0004-6361/202243782}

\bibitem[{{Gaia Collaboration} {et~al.}(2023{\natexlab{b}}){Gaia Collaboration}, {Vallenari}, {Brown}, {Prusti}, {de Bruijne}, {Arenou}, {Babusiaux}, {Biermann}, {Creevey}, {Ducourant}, {Evans}, {Eyer}, {Guerra}, {Hutton}, {Jordi}, {Klioner}, {Lammers}, {Lindegren}, {Luri}, {Mignard}, {Panem}, {Pourbaix}, {Randich}, {Sartoretti}, {Soubiran}, {Tanga}, {Walton}, {Bailer-Jones}, {Bastian}, {Drimmel}, {Jansen}, {Katz}, {Lattanzi}, {van Leeuwen}, {Bakker}, {Cacciari}, {Casta{\~n}eda}, {De Angeli}, {Fabricius}, {Fouesneau}, {Fr{\'e}mat}, {Galluccio}, {Guerrier}, {Heiter}, {Masana}, {Messineo}, {Mowlavi}, {Nicolas}, {Nienartowicz}, {Pailler}, {Panuzzo}, {Riclet}, {Roux}, {Seabroke}, {Sordo}, {Th{\'e}venin}, {Gracia-Abril}, {Portell}, {Teyssier}, {Altmann}, {Andrae}, {Audard}, {Bellas-Velidis}, {Benson}, {Berthier}, {Blomme}, {Burgess}, {Busonero}, {Busso}, {C{\'a}novas}, {Carry}, {Cellino}, {Cheek}, {Clementini}, {Damerdji}, {Davidson}, {de Teodoro}, {Nu{\~n}ez Campos}, {Delchambre}, {Dell'Oro}, {Esquej},
  {Fern{\'a}ndez-Hern{\'a}ndez}, {Fraile}, {Garabato}, {Garc{\'\i}a-Lario}, {Gosset}, {Haigron}, {Halbwachs}, {Hambly}, {Harrison}, {Hern{\'a}ndez}, {Hestroffer}, {Hodgkin}, {Holl}, {Jan{\ss}en}, {Jevardat de Fombelle}, {Jordan}, {Krone-Martins}, {Lanzafame}, {L{\"o}ffler}, {Marchal}, {Marrese}, {Moitinho}, {Muinonen}, {Osborne}, {Pancino}, {Pauwels}, {Recio-Blanco}, {Reyl{\'e}}, {Riello}, {Rimoldini}, {Roegiers}, {Rybizki}, {Sarro}, {Siopis}, {Smith}, {Sozzetti}, {Utrilla}, {van Leeuwen}, {Abbas}, {{\'A}brah{\'a}m}, {Abreu Aramburu}, {Aerts}, {Aguado}, {Ajaj}, {Aldea-Montero}, {Altavilla}, {{\'A}lvarez}, {Alves}, {Anders}, {Anderson}, {Anglada Varela}, {Antoja}, {Baines}, {Baker}, {Balaguer-N{\'u}{\~n}ez}, {Balbinot}, {Balog}, {Barache}, {Barbato}, {Barros}, {Barstow}, {Bartolom{\'e}}, {Bassilana}, {Bauchet}, {Becciani}, {Bellazzini}, {Berihuete}, {Bernet}, {Bertone}, {Bianchi}, {Binnenfeld}, {Blanco-Cuaresma}, {Blazere}, {Boch}, {Bombrun}, {Bossini}, {Bouquillon}, {Bragaglia}, {Bramante}, {Breedt},
  {Bressan}, {Brouillet}, {Brugaletta}, {Bucciarelli}, {Burlacu}, {Butkevich}, {Buzzi}, {Caffau}, {Cancelliere}, {Cantat-Gaudin}, {Carballo}, {Carlucci}, {Carnerero}, {Carrasco}, {Casamiquela}, {Castellani}, {Castro-Ginard}, {Chaoul}, {Charlot}, {Chemin}, {Chiaramida}, {Chiavassa}, {Chornay}, {Comoretto}, {Contursi}, {Cooper}, {Cornez}, {Cowell}, {Crifo}, {Cropper}, {Crosta}, {Crowley}, {Dafonte}, {Dapergolas}, {David}, {David}, {de Laverny}, {De Luise}, {De March}, {De Ridder}, {de Souza}, {de Torres}, {del Peloso}, {del Pozo}, {Delbo}, {Delgado}, {Delisle}, {Demouchy}, {Dharmawardena}, {Di Matteo}, {Diakite}, {Diener}, {Distefano}, {Dolding}, {Edvardsson}, {Enke}, {Fabre}, {Fabrizio}, {Faigler}, {Fedorets}, {Fernique}, {Fienga}, {Figueras}, {Fournier}, {Fouron}, {Fragkoudi}, {Gai}, {Garcia-Gutierrez}, {Garcia-Reinaldos}, {Garc{\'\i}a-Torres}, {Garofalo}, {Gavel}, {Gavras}, {Gerlach}, {Geyer}, {Giacobbe}, {Gilmore}, {Girona}, {Giuffrida}, {Gomel}, {Gomez}, {Gonz{\'a}lez-N{\'u}{\~n}ez},
  {Gonz{\'a}lez-Santamar{\'\i}a}, {Gonz{\'a}lez-Vidal}, {Granvik}, {Guillout}, {Guiraud}, {Guti{\'e}rrez-S{\'a}nchez}, {Guy}, {Hatzidimitriou}, {Hauser}, {Haywood}, {Helmer}, {Helmi}, {Sarmiento}, {Hidalgo}, {Hilger}, {H{\l}adczuk}, {Hobbs}, {Holland}, {Huckle}, {Jardine}, {Jasniewicz}, {Jean-Antoine Piccolo}, {Jim{\'e}nez-Arranz}, {Jorissen}, {Juaristi Campillo}, {Julbe}, {Karbevska}, {Kervella}, {Khanna}, {Kontizas}, {Kordopatis}, {Korn}, {K{\'o}sp{\'a}l}, {Kostrzewa-Rutkowska}, {Kruszy{\'n}ska}, {Kun}, {Laizeau}, {Lambert}, {Lanza}, {Lasne}, {Le Campion}, {Lebreton}, {Lebzelter}, {Leccia}, {Leclerc}, {Lecoeur-Taibi}, {Liao}, {Licata}, {Lindstr{\o}m}, {Lister}, {Livanou}, {Lobel}, {Lorca}, {Loup}, {Madrero Pardo}, {Magdaleno Romeo}, {Managau}, {Mann}, {Manteiga}, {Marchant}, {Marconi}, {Marcos}, {Marcos Santos}, {Mar{\'\i}n Pina}, {Marinoni}, {Marocco}, {Marshall}, {Martin Polo}, {Mart{\'\i}n-Fleitas}, {Marton}, {Mary}, {Masip}, {Massari}, {Mastrobuono-Battisti}, {Mazeh}, {McMillan}, {Messina}, {Michalik},
  {Millar}, {Mints}, {Molina}, {Molinaro}, {Moln{\'a}r}, {Monari}, {Mongui{\'o}}, {Montegriffo}, {Montero}, {Mor}, {Mora}, {Morbidelli}, {Morel}, {Morris}, {Muraveva}, {Murphy}, {Musella}, {Nagy}, {Noval}, {Oca{\~n}a}, {Ogden}, {Ordenovic}, {Osinde}, {Pagani}, {Pagano}, {Palaversa}, {Palicio}, {Pallas-Quintela}, {Panahi}, {Payne-Wardenaar}, {Pe{\~n}alosa Esteller}, {Penttil{\"a}}, {Pichon}, {Piersimoni}, {Pineau}, {Plachy}, {Plum}, {Poggio}, {Pr{\v{s}}a}, {Pulone}, {Racero}, {Ragaini}, {Rainer}, {Raiteri}, {Rambaux}, {Ramos}, {Ramos-Lerate}, {Re Fiorentin}, {Regibo}, {Richards}, {Rios Diaz}, {Ripepi}, {Riva}, {Rix}, {Rixon}, {Robichon}, {Robin}, {Robin}, {Roelens}, {Rogues}, {Rohrbasser}, {Romero-G{\'o}mez}, {Rowell}, {Royer}, {Ruz Mieres}, {Rybicki}, {Sadowski}, {S{\'a}ez N{\'u}{\~n}ez}, {Sagrist{\`a} Sell{\'e}s}, {Sahlmann}, {Salguero}, {Samaras}, {Sanchez Gimenez}, {Sanna}, {Santove{\~n}a}, {Sarasso}, {Schultheis}, {Sciacca}, {Segol}, {Segovia}, {S{\'e}gransan}, {Semeux}, {Shahaf}, {Siddiqui}, {Siebert},
  {Siltala}, {Silvelo}, {Slezak}, {Slezak}, {Smart}, {Snaith}, {Solano}, {Solitro}, {Souami}, {Souchay}, {Spagna}, {Spina}, {Spoto}, {Steele}, {Steidelm{\"u}ller}, {Stephenson}, {S{\"u}veges}, {Surdej}, {Szabados}, {Szegedi-Elek}, {Taris}, {Taylor}, {Teixeira}, {Tolomei}, {Tonello}, {Torra}, {Torra}, {Torralba Elipe}, {Trabucchi}, {Tsounis}, {Turon}, {Ulla}, {Unger}, {Vaillant}, {van Dillen}, {van Reeven}, {Vanel}, {Vecchiato}, {Viala}, {Vicente}, {Voutsinas}, {Weiler}, {Wevers}, {Wyrzykowski}, {Yoldas}, {Yvard}, {Zhao}, {Zorec}, {Zucker}, \& {Zwitter}}]{2023A&A...674A...1G}
{Gaia Collaboration}, {Vallenari}, A., {Brown}, A.~G.~A., {et~al.} 2023{\natexlab{b}}, \aap, 674, A1, \dodoi{10.1051/0004-6361/202243940}

\bibitem[{{Gaia Collaboration} {et~al.}(2023{\natexlab{c}}){Gaia Collaboration}, {Drimmel}, {Romero-G{\'o}mez}, {Chemin}, {Ramos}, {Poggio}, {Ripepi}, {Andrae}, {Blomme}, {Cantat-Gaudin}, {Castro-Ginard}, {Clementini}, {Figueras}, {Fouesneau}, {Fr{\'e}mat}, {Jardine}, {Khanna}, {Lobel}, {Marshall}, {Muraveva}, {Brown}, {Vallenari}, {Prusti}, {de Bruijne}, {Arenou}, {Babusiaux}, {Biermann}, {Creevey}, {Ducourant}, {Evans}, {Eyer}, {Guerra}, {Hutton}, {Jordi}, {Klioner}, {Lammers}, {Lindegren}, {Luri}, {Mignard}, {Panem}, {Pourbaix}, {Randich}, {Sartoretti}, {Soubiran}, {Tanga}, {Walton}, {Bailer-Jones}, {Bastian}, {Jansen}, {Katz}, {Lattanzi}, {van Leeuwen}, {Bakker}, {Cacciari}, {Casta{\~n}eda}, {De Angeli}, {Fabricius}, {Galluccio}, {Guerrier}, {Heiter}, {Masana}, {Messineo}, {Mowlavi}, {Nicolas}, {Nienartowicz}, {Pailler}, {Panuzzo}, {Riclet}, {Roux}, {Seabroke}, {Sordo}, {Th{\'e}venin}, {Gracia-Abril}, {Portell}, {Teyssier}, {Altmann}, {Audard}, {Bellas-Velidis}, {Benson}, {Berthier}, {Burgess},
  {Busonero}, {Busso}, {C{\'a}novas}, {Carry}, {Cellino}, {Cheek}, {Damerdji}, {Davidson}, {de Teodoro}, {Nu{\~n}ez Campos}, {Delchambre}, {Dell'Oro}, {Esquej}, {Fern{\'a}ndez-Hern{\'a}ndez}, {Fraile}, {Garabato}, {Garc{\'\i}a-Lario}, {Gosset}, {Haigron}, {Halbwachs}, {Hambly}, {Harrison}, {Hern{\'a}ndez}, {Hestroffer}, {Hodgkin}, {Holl}, {Jan{\ss}en}, {Jevardat de Fombelle}, {Jordan}, {Krone-Martins}, {Lanzafame}, {L{\"o}ffler}, {Marchal}, {Marrese}, {Moitinho}, {Muinonen}, {Osborne}, {Pancino}, {Pauwels}, {Recio-Blanco}, {Reyl{\'e}}, {Riello}, {Rimoldini}, {Roegiers}, {Rybizki}, {Sarro}, {Siopis}, {Smith}, {Sozzetti}, {Utrilla}, {van Leeuwen}, {Abbas}, {{\'A}brah{\'a}m}, {Abreu Aramburu}, {Aerts}, {Aguado}, {Ajaj}, {Aldea-Montero}, {Altavilla}, {{\'A}lvarez}, {Alves}, {Anders}, {Anderson}, {Anglada Varela}, {Antoja}, {Baines}, {Baker}, {Balaguer-N{\'u}{\~n}ez}, {Balbinot}, {Balog}, {Barache}, {Barbato}, {Barros}, {Barstow}, {Bartolom{\'e}}, {Bassilana}, {Bauchet}, {Becciani}, {Bellazzini}, {Berihuete},
  {Bernet}, {Bertone}, {Bianchi}, {Binnenfeld}, {Blanco-Cuaresma}, {Boch}, {Bombrun}, {Bossini}, {Bouquillon}, {Bragaglia}, {Bramante}, {Breedt}, {Bressan}, {Brouillet}, {Brugaletta}, {Bucciarelli}, {Burlacu}, {Butkevich}, {Buzzi}, {Caffau}, {Cancelliere}, {Carballo}, {Carlucci}, {Carnerero}, {Carrasco}, {Casamiquela}, {Castellani}, {Chaoul}, {Charlot}, {Chiaramida}, {Chiavassa}, {Chornay}, {Comoretto}, {Contursi}, {Cooper}, {Cornez}, {Cowell}, {Crifo}, {Cropper}, {Crosta}, {Crowley}, {Dafonte}, {Dapergolas}, {David}, {de Laverny}, {De Luise}, {De March}, {De Ridder}, {de Souza}, {de Torres}, {del Peloso}, {del Pozo}, {Delbo}, {Delgado}, {Delisle}, {Demouchy}, {Dharmawardena}, {Di Matteo}, {Diakite}, {Diener}, {Distefano}, {Dolding}, {Enke}, {Fabre}, {Fabrizio}, {Faigler}, {Fedorets}, {Fernique}, {Fournier}, {Fouron}, {Fragkoudi}, {Gai}, {Garcia-Gutierrez}, {Garcia-Reinaldos}, {Garc{\'\i}a-Torres}, {Garofalo}, {Gavel}, {Gavras}, {Gerlach}, {Geyer}, {Giacobbe}, {Gilmore}, {Girona}, {Giuffrida}, {Gomel},
  {Gomez}, {Gonz{\'a}lez-N{\'u}{\~n}ez}, {Gonz{\'a}lez-Santamar{\'\i}a}, {Gonz{\'a}lez-Vidal}, {Granvik}, {Guillout}, {Guiraud}, {Guti{\'e}rrez-S{\'a}nchez}, {Guy}, {Hatzidimitriou}, {Hauser}, {Haywood}, {Helmer}, {Helmi}, {Sarmiento}, {Hidalgo}, {H{\l}adczuk}, {Hobbs}, {Holland}, {Huckle}, {Jasniewicz}, {Jean-Antoine Piccolo}, {Jim{\'e}nez-Arranz}, {Juaristi Campillo}, {Julbe}, {Karbevska}, {Kervella}, {Kordopatis}, {Korn}, {K{\'o}sp{\'a}l}, {Kostrzewa-Rutkowska}, {Kruszy{\'n}ska}, {Kun}, {Laizeau}, {Lambert}, {Lanza}, {Lasne}, {Le Campion}, {Lebreton}, {Lebzelter}, {Leccia}, {Leclerc}, {Lecoeur-Taibi}, {Liao}, {Licata}, {Lindstr{\o}m}, {Lister}, {Livanou}, {Lorca}, {Loup}, {Madrero Pardo}, {Magdaleno Romeo}, {Managau}, {Mann}, {Manteiga}, {Marchant}, {Marconi}, {Marcos}, {Marcos Santos}, {Mar{\'\i}n Pina}, {Marinoni}, {Marocco}, {Martin Polo}, {Mart{\'\i}n-Fleitas}, {Marton}, {Mary}, {Masip}, {Massari}, {Mastrobuono-Battisti}, {Mazeh}, {McMillan}, {Messina}, {Michalik}, {Millar}, {Mints}, {Molina},
  {Molinaro}, {Moln{\'a}r}, {Monari}, {Mongui{\'o}}, {Montegriffo}, {Montero}, {Mor}, {Mora}, {Morbidelli}, {Morel}, {Morris}, {Murphy}, {Musella}, {Nagy}, {Noval}, {Oca{\~n}a}, {Ogden}, {Ordenovic}, {Osinde}, {Pagani}, {Pagano}, {Palaversa}, {Palicio}, {Pallas-Quintela}, {Panahi}, {Payne-Wardenaar}, {Pe{\~n}alosa Esteller}, {Penttil{\"a}}, {Pichon}, {Piersimoni}, {Pineau}, {Plachy}, {Plum}, {Pr{\v{s}}a}, {Pulone}, {Racero}, {Ragaini}, {Rainer}, {Raiteri}, {Ramos-Lerate}, {Re Fiorentin}, {Regibo}, {Richards}, {Rios Diaz}, {Riva}, {Rix}, {Rixon}, {Robichon}, {Robin}, {Robin}, {Roelens}, {Rogues}, {Rohrbasser}, {Rowell}, {Royer}, {Ruz Mieres}, {Rybicki}, {Sadowski}, {S{\'a}ez N{\'u}{\~n}ez}, {Sagrist{\`a} Sell{\'e}s}, {Sahlmann}, {Salguero}, {Samaras}, {Sanchez Gimenez}, {Sanna}, {Santove{\~n}a}, {Sarasso}, {Schultheis}, {Sciacca}, {Segol}, {Segovia}, {S{\'e}gransan}, {Semeux}, {Shahaf}, {Siddiqui}, {Siebert}, {Siltala}, {Silvelo}, {Slezak}, {Slezak}, {Smart}, {Snaith}, {Solano}, {Solitro}, {Souami}, {Souchay},
  {Spagna}, {Spina}, {Spoto}, {Steele}, {Steidelm{\"u}ller}, {Stephenson}, {S{\"u}veges}, {Surdej}, {Szabados}, {Szegedi-Elek}, {Taris}, {Taylor}, {Teixeira}, {Tolomei}, {Tonello}, {Torra}, {Torra}, {Torralba Elipe}, {Trabucchi}, {Tsounis}, {Turon}, {Ulla}, {Unger}, {Vaillant}, {van Dillen}, {van Reeven}, {Vanel}, {Vecchiato}, {Viala}, {Vicente}, {Voutsinas}, {Weiler}, {Wevers}, {Wyrzykowski}, {Yoldas}, {Yvard}, {Zhao}, {Zorec}, {Zucker}, \& {Zwitter}}]{2023A&A...674A..37G}
{Gaia Collaboration}, {Drimmel}, R., {Romero-G{\'o}mez}, M., {et~al.} 2023{\natexlab{c}}, \aap, 674, A37, \dodoi{10.1051/0004-6361/202243797}

\bibitem[{{Green} {et~al.}(2019){Green}, {Schlafly}, {Zucker}, {Speagle}, \& {Finkbeiner}}]{2019ApJ...887...93G}
{Green}, G.~M., {Schlafly}, E., {Zucker}, C., {Speagle}, J.~S., \& {Finkbeiner}, D. 2019, \apj, 887, 93, \dodoi{10.3847/1538-4357/ab5362}

\bibitem[{{Guo} {et~al.}(2017){Guo}, {Gies}, {Matson}, {Garc{\'\i}a Hern{\'a}ndez}, {Han}, \& {Chen}}]{2017ApJ...837..114G}
{Guo}, Z., {Gies}, D.~R., {Matson}, R.~A., {et~al.} 2017, \apj, 837, 114, \dodoi{10.3847/1538-4357/aa61a4}

\bibitem[{{Henden} \& {Munari}(2014)}]{2014CoSka..43..518H}
{Henden}, A., \& {Munari}, U. 2014, Contributions of the Astronomical Observatory Skalnate Pleso, 43, 518

\bibitem[{{Hong} {et~al.}(2021){Hong}, {Lee}, {Koo}, {Park}, {Rittipruk}, {Kim}, {Kanjanasakul}, \& {Han}}]{2021AJ....161..137H}
{Hong}, K., {Lee}, J.~W., {Koo}, J.-R., {et~al.} 2021, \aj, 161, 137, \dodoi{10.3847/1538-3881/abdd39}

\bibitem[{{Howard} {et~al.}(2022){Howard}, {Davenport}, \& {Covey}}]{2022RNAAS...6...96H}
{Howard}, E.~L., {Davenport}, J. R.~A., \& {Covey}, K.~R. 2022, Research Notes of the American Astronomical Society, 6, 96, \dodoi{10.3847/2515-5172/ac6e42}

\bibitem[{{Huang} {et~al.}(2020{\natexlab{a}}){Huang}, {Vanderburg}, {P{\'a}l}, {Sha}, {Yu}, {Fong}, {Fausnaugh}, {Shporer}, {Guerrero}, {Vanderspek}, \& {Ricker}}]{2020RNAAS...4..204H}
{Huang}, C.~X., {Vanderburg}, A., {P{\'a}l}, A., {et~al.} 2020{\natexlab{a}}, Research Notes of the American Astronomical Society, 4, 204, \dodoi{10.3847/2515-5172/abca2e}

\bibitem[{{Huang} {et~al.}(2020{\natexlab{b}}){Huang}, {Vanderburg}, {P{\'a}l}, {Sha}, {Yu}, {Fong}, {Fausnaugh}, {Shporer}, {Guerrero}, {Vanderspek}, \& {Ricker}}]{2020RNAAS...4..206H}
---. 2020{\natexlab{b}}, Research Notes of the American Astronomical Society, 4, 206, \dodoi{10.3847/2515-5172/abca2d}

\bibitem[{{IJspeert} {et~al.}(2021){IJspeert}, {Tkachenko}, {Johnston}, {Garcia}, {De Ridder}, {Van Reeth}, \& {Aerts}}]{2021AA...652A.120I}
{IJspeert}, L.~W., {Tkachenko}, A., {Johnston}, C., {et~al.} 2021, \aap, 652, A120, \dodoi{10.1051/0004-6361/202141489}

\bibitem[{{Jones} {et~al.}(2020){Jones}, {Conroy}, {Horvat}, {Giammarco}, {Kochoska}, {Pablo}, {Brown}, {Sowicka}, \& {Pr{\v{s}}a}}]{2020ApJS..247...63J}
{Jones}, D., {Conroy}, K.~E., {Horvat}, M., {et~al.} 2020, \apjs, 247, 63, \dodoi{10.3847/1538-4365/ab7927}

\bibitem[{{Kim} {et~al.}(2021){Kim}, {Lee}, {Lee}, {Lee}, {Lee}, {Hong}, {Cha}, {Kim}, \& {Park}}]{2021AJ....162..212K}
{Kim}, S.-L., {Lee}, J.~W., {Lee}, C.-U., {et~al.} 2021, \aj, 162, 212, \dodoi{10.3847/1538-3881/ac23de}

\bibitem[{{Kunimoto} {et~al.}(2022){Kunimoto}, {Tey}, {Fong}, {Hesse}, {Shporer}, {Fausnaugh}, {Vanderspek}, \& {Ricker}}]{2022RNAAS...6..236K}
{Kunimoto}, M., {Tey}, E., {Fong}, W., {et~al.} 2022, Research Notes of the American Astronomical Society, 6, 236, \dodoi{10.3847/2515-5172/aca158}

\bibitem[{{Kunimoto} {et~al.}(2021){Kunimoto}, {Huang}, {Tey}, {Fong}, {Hesse}, {Shporer}, {Guerrero}, {Fausnaugh}, {Vanderspek}, \& {Ricker}}]{2021RNAAS...5..234K}
{Kunimoto}, M., {Huang}, C., {Tey}, E., {et~al.} 2021, Research Notes of the American Astronomical Society, 5, 234, \dodoi{10.3847/2515-5172/ac2ef0}

\bibitem[{{Kurucz}(1979)}]{1979ApJS...40....1K}
{Kurucz}, R.~L. 1979, \apjs, 40, 1, \dodoi{10.1086/190589}

\bibitem[{{Lee} {et~al.}(2022{\natexlab{a}}){Lee}, {Hong}, {Kim}, \& {Park}}]{2022MNRAS.515.4702L}
{Lee}, J.~W., {Hong}, K., {Kim}, H.-Y., \& {Park}, J.-H. 2022{\natexlab{a}}, \mnras, 515, 4702, \dodoi{10.1093/mnras/stac2151}

\bibitem[{{Lee} {et~al.}(2022{\natexlab{b}}){Lee}, {Hong}, \& {Park}}]{2022MNRAS.511..654L}
{Lee}, J.~W., {Hong}, K., \& {Park}, J.-H. 2022{\natexlab{b}}, \mnras, 511, 654, \dodoi{10.1093/mnras/stac075}

\bibitem[{{Lee} {et~al.}(2020){Lee}, {Koo}, {Hong}, \& {Park}}]{2020AJ....160...49L}
{Lee}, J.~W., {Koo}, J.-R., {Hong}, K., \& {Park}, J.-H. 2020, \aj, 160, 49, \dodoi{10.3847/1538-3881/ab9621}

\bibitem[{{Li} {et~al.}(2019){Li}, {Chen}, {Chen}, \& {Han}}]{2019ApJ...871..148L}
{Li}, Z., {Chen}, X., {Chen}, H.-L., \& {Han}, Z. 2019, \apj, 871, 148, \dodoi{10.3847/1538-4357/aaf9a1}

\bibitem[{{Lightkurve Collaboration} {et~al.}(2018){Lightkurve Collaboration}, {Cardoso}, {Hedges}, {Gully-Santiago}, {Saunders}, {Cody}, {Barclay}, {Hall}, {Sagear}, {Turtelboom}, {Zhang}, {Tzanidakis}, {Mighell}, {Coughlin}, {Bell}, {Berta-Thompson}, {Williams}, {Dotson}, \& {Barentsen}}]{2018ascl}
{Lightkurve Collaboration}, {Cardoso}, J. V. d.~M., {Hedges}, C., {et~al.} 2018, {Lightkurve: Kepler and TESS time series analysis in Python}, Astrophysics Source Code Library, record ascl:1812.013

\bibitem[{{Lin} {et~al.}(2011){Lin}, {Rappaport}, {Podsiadlowski}, {Nelson}, {Paxton}, \& {Todorov}}]{2011ApJ...732...70L}
{Lin}, J., {Rappaport}, S., {Podsiadlowski}, P., {et~al.} 2011, \apj, 732, 70, \dodoi{10.1088/0004-637X/732/2/70}

\bibitem[{{Liu} {et~al.}(2020){Liu}, {Fu}, {Shi}, {Wu}, {Han}, {Chen}, {Dong}, {Zhao}, {Chen}, {Zhang}, {Bai}, {Chen}, {Cui}, {Du}, {Hsia}, {Jiang}, {Hou}, {Hou}, {Li}, {Li}, {Li}, {Liu}, {Liu}, {Luo}, {Ren}, {Tian}, {Tian}, {Wang}, {Wu}, {Xie}, {Yan}, {Yang}, {Yu}, {Zhang}, {Zhang}, {Zhang}, {Zhang}, {Zhao}, {Zhong}, {Zong}, \& {Zuo}}]{2020arXiv200507210L}
{Liu}, C., {Fu}, J., {Shi}, J., {et~al.} 2020, arXiv e-prints, arXiv:2005.07210.
\newblock \doarXiv{2005.07210}

\bibitem[{{Lucy}(1967)}]{1967ZA.....65...89L}
{Lucy}, L.~B. 1967, \zap, 65, 89

\bibitem[{{Luo} {et~al.}(2015){Luo}, {Zhao}, {Zhao}, {Deng}, {Liu}, {Jing}, {Wang}, {Zhang}, {Shi}, {Cui}, {Chu}, {Li}, {Bai}, {Wu}, {Cai}, {Cao}, {Cao}, {Carlin}, {Chen}, {Chen}, {Chen}, {Chen}, {Chen}, {Chen}, {Chen}, {Christlieb}, {Chu}, {Cui}, {Dong}, {Du}, {Fan}, {Feng}, {Fu}, {Gao}, {Gong}, {Gu}, {Guo}, {Han}, {He}, {Hou}, {Hou}, {Hou}, {Hu}, {Hu}, {Hu}, {Huo}, {Jia}, {Jiang}, {Jiang}, {Jiang}, {Jin}, {Kong}, {Kong}, {Lei}, {Li}, {Li}, {Li}, {Li}, {Li}, {Li}, {Li}, {Li}, {Li}, {Li}, {Li}, {Li}, {Liang}, {Lin}, {Liu}, {Liu}, {Liu}, {Liu}, {Lu}, {Luo}, {Mao}, {Newberg}, {Ni}, {Qi}, {Qi}, {Shen}, {Shi}, {Song}, {Song}, {Su}, {Su}, {Tang}, {Tao}, {Tian}, {Wang}, {Wang}, {Wang}, {Wang}, {Wang}, {Wang}, {Wang}, {Wang}, {Wang}, {Wang}, {Wang}, {Wang}, {Wang}, {Wang}, {Wang}, {Wang}, {Wang}, {Wang}, {Wang}, {Wang}, {Wei}, {Wei}, {Wu}, {Wu}, {Wu}, {Wu}, {Xing}, {Xu}, {Xu}, {Xu}, {Yan}, {Yang}, {Yang}, {Yang}, {Yang}, {Yao}, {Yu}, {Yuan}, {Yuan}, {Yuan}, {Yuan}, {Zhai}, {Zhang}, {Zhang}, {Zhang}, {Zhang},
  {Zhang}, {Zhang}, {Zhang}, {Zhang}, {Zhao}, {Zhou}, {Zhou}, {Zhu}, {Zhu}, {Zou}, \& {Zuo}}]{2015RAA....15.1095L}
{Luo}, A.~L., {Zhao}, Y.-H., {Zhao}, G., {et~al.} 2015, Research in Astronomy and Astrophysics, 15, 1095, \dodoi{10.1088/1674-4527/15/8/002}

\bibitem[{{Luo}(2020)}]{2020NewA...7801363L}
{Luo}, Y. 2020, \na, 78, 101363, \dodoi{10.1016/j.newast.2020.101363}

\bibitem[{{Matijevi{\v{c}}} {et~al.}(2012){Matijevi{\v{c}}}, {Pr{\v{s}}a}, {Orosz}, {Welsh}, {Bloemen}, \& {Barclay}}]{2012AJ....143..123M}
{Matijevi{\v{c}}}, G., {Pr{\v{s}}a}, A., {Orosz}, J.~A., {et~al.} 2012, \aj, 143, 123, \dodoi{10.1088/0004-6256/143/5/123}

\bibitem[{{Maxted} {et~al.}(2011){Maxted}, {Anderson}, {Burleigh}, {Collier Cameron}, {Heber}, {G{\"a}nsicke}, {Geier}, {Kupfer}, {Marsh}, {Nelemans}, {O'Toole}, {{\O}stensen}, {Smalley}, \& {West}}]{2011MNRAS.418.1156M}
{Maxted}, P.~F.~L., {Anderson}, D.~R., {Burleigh}, M.~R., {et~al.} 2011, \mnras, 418, 1156, \dodoi{10.1111/j.1365-2966.2011.19567.x}

\bibitem[{{Maxted} {et~al.}(2014){Maxted}, {Bloemen}, {Heber}, {Geier}, {Wheatley}, {Marsh}, {Breedt}, {Sebastian}, {Faillace}, {Owen}, {Pulley}, {Smith}, {Kolb}, {Haswell}, {Southworth}, {Anderson}, {Smalley}, {Collier Cameron}, {Hebb}, {Simpson}, {West}, {Bochinski}, {Busuttil}, \& {Hadigal}}]{2014MNRAS.437.1681M}
{Maxted}, P.~F.~L., {Bloemen}, S., {Heber}, U., {et~al.} 2014, \mnras, 437, 1681, \dodoi{10.1093/mnras/stt2007}

\bibitem[{{Onken} {et~al.}(2019){Onken}, {Wolf}, {Bessell}, {Chang}, {Da Costa}, {Luvaul}, {Mackey}, {Schmidt}, \& {Shao}}]{2019PASA...36...33O}
{Onken}, C.~A., {Wolf}, C., {Bessell}, M.~S., {et~al.} 2019, \pasa, 36, e033, \dodoi{10.1017/pasa.2019.27}

\bibitem[{{Paxton} {et~al.}(2011){Paxton}, {Bildsten}, {Dotter}, {Herwig}, {Lesaffre}, \& {Timmes}}]{2011ApJS..192....3P}
{Paxton}, B., {Bildsten}, L., {Dotter}, A., {et~al.} 2011, \apjs, 192, 3, \dodoi{10.1088/0067-0049/192/1/3}

\bibitem[{{Paxton} {et~al.}(2013){Paxton}, {Cantiello}, {Arras}, {Bildsten}, {Brown}, {Dotter}, {Mankovich}, {Montgomery}, {Stello}, {Timmes}, \& {Townsend}}]{2013ApJS..208....4P}
{Paxton}, B., {Cantiello}, M., {Arras}, P., {et~al.} 2013, \apjs, 208, 4, \dodoi{10.1088/0067-0049/208/1/4}

\bibitem[{{Paxton} {et~al.}(2015){Paxton}, {Marchant}, {Schwab}, {Bauer}, {Bildsten}, {Cantiello}, {Dessart}, {Farmer}, {Hu}, {Langer}, {Townsend}, {Townsley}, \& {Timmes}}]{2015ApJS..220...15P}
{Paxton}, B., {Marchant}, P., {Schwab}, J., {et~al.} 2015, \apjs, 220, 15, \dodoi{10.1088/0067-0049/220/1/15}

\bibitem[{{Peng} {et~al.}(2024){Peng}, {Wang}, \& {Ren}}]{2024NewA..10702153P}
{Peng}, Y., {Wang}, K., \& {Ren}, A. 2024, \na, 107, 102153, \dodoi{10.1016/j.newast.2023.102153}

\bibitem[{{Pr{\v{s}}a} \& {Zwitter}(2005)}]{2005ApJ...628..426P}
{Pr{\v{s}}a}, A., \& {Zwitter}, T. 2005, \apj, 628, 426, \dodoi{10.1086/430591}

\bibitem[{{Pr{\v{s}}a} {et~al.}(2016){Pr{\v{s}}a}, {Conroy}, {Horvat}, {Pablo}, {Kochoska}, {Bloemen}, {Giammarco}, {Hambleton}, \& {Degroote}}]{2016ApJS..227...29P}
{Pr{\v{s}}a}, A., {Conroy}, K.~E., {Horvat}, M., {et~al.} 2016, \apjs, 227, 29, \dodoi{10.3847/1538-4365/227/2/29}

\bibitem[{{Pr{\v{s}}a} {et~al.}(2022){Pr{\v{s}}a}, {Kochoska}, {Conroy}, {Eisner}, {Hey}, {IJspeert}, {Kruse}, {Fleming}, {Johnston}, {Kristiansen}, {LaCourse}, {Mortensen}, {Pepper}, {Stassun}, {Torres}, {Abdul-Masih}, {Chakraborty}, {Gagliano}, {Guo}, {Hambleton}, {Hong}, {Jacobs}, {Jones}, {Kostov}, {Lee}, {Omohundro}, {Orosz}, {Page}, {Powell}, {Rappaport}, {Reed}, {Schnittman}, {Schwengeler}, {Shporer}, {Terentev}, {Vanderburg}, {Welsh}, {Caldwell}, {Doty}, {Jenkins}, {Latham}, {Ricker}, {Seager}, {Schlieder}, {Shiao}, {Vanderspek}, \& {Winn}}]{2022ApJS..258...16P}
{Pr{\v{s}}a}, A., {Kochoska}, A., {Conroy}, K.~E., {et~al.} 2022, \apjs, 258, 16, \dodoi{10.3847/1538-4365/ac324a}

\bibitem[{{Rappaport} {et~al.}(2015){Rappaport}, {Nelson}, {Levine}, {Sanchis-Ojeda}, {Gandolfi}, {Nowak}, {Palle}, \& {Prsa}}]{2015ApJ...803...82R}
{Rappaport}, S., {Nelson}, L., {Levine}, A., {et~al.} 2015, \apj, 803, 82, \dodoi{10.1088/0004-637X/803/2/82}

\bibitem[{{Ruci{\'n}ski}(1969)}]{1969AcA....19..245R}
{Ruci{\'n}ski}, S.~M. 1969, \actaa, 19, 245

\bibitem[{{Schlegel} {et~al.}(1998){Schlegel}, {Finkbeiner}, \& {Davis}}]{1998ApJ...500..525S}
{Schlegel}, D.~J., {Finkbeiner}, D.~P., \& {Davis}, M. 1998, \apj, 500, 525, \dodoi{10.1086/305772}

\bibitem[{{Smith} {et~al.}(2012){Smith}, {Stumpe}, {Van Cleve}, {Jenkins}, {Barclay}, {Fanelli}, {Girouard}, {Kolodziejczak}, {McCauliff}, {Morris}, \& {Twicken}}]{2012PASP..124.1000S}
{Smith}, J.~C., {Stumpe}, M.~C., {Van Cleve}, J.~E., {et~al.} 2012, \pasp, 124, 1000, \dodoi{10.1086/667697}

\bibitem[{{Stumpe} {et~al.}(2014){Stumpe}, {Smith}, {Catanzarite}, {Van Cleve}, {Jenkins}, {Twicken}, \& {Girouard}}]{2014PASP..126..100S}
{Stumpe}, M.~C., {Smith}, J.~C., {Catanzarite}, J.~H., {et~al.} 2014, \pasp, 126, 100, \dodoi{10.1086/674989}

\bibitem[{{van Kerkwijk} {et~al.}(2010){van Kerkwijk}, {Rappaport}, {Breton}, {Justham}, {Podsiadlowski}, \& {Han}}]{2010ApJ...715...51V}
{van Kerkwijk}, M.~H., {Rappaport}, S.~A., {Breton}, R.~P., {et~al.} 2010, \apj, 715, 51, \dodoi{10.1088/0004-637X/715/1/51}

\bibitem[{{van Roestel} {et~al.}(2018){van Roestel}, {Kupfer}, {Ruiz-Carmona}, {Groot}, {Prince}, {Burdge}, {Laher}, {Shupe}, \& {Bellm}}]{2018MNRAS.475.2560V}
{van Roestel}, J., {Kupfer}, T., {Ruiz-Carmona}, R., {et~al.} 2018, \mnras, 475, 2560, \dodoi{10.1093/mnras/stx3291}

\bibitem[{{von Zeipel}(1924)}]{1924MNRAS..84..665V}
{von Zeipel}, H. 1924, \mnras, 84, 665, \dodoi{10.1093/mnras/84.9.665}

\bibitem[{{Wang} {et~al.}(2020{\natexlab{a}}){Wang}, {Zhang}, \& {Dai}}]{2020ApJ...888...49W}
{Wang}, K., {Zhang}, X., \& {Dai}, M. 2020{\natexlab{a}}, \apj, 888, 49, \dodoi{10.3847/1538-4357/ab584c}

\bibitem[{{Wang} {et~al.}(2020{\natexlab{b}}){Wang}, {Gies}, {Lester}, {Guo}, {Matson}, {Peters}, {Dhillon}, {Butterley}, {Littlefair}, {Wilson}, \& {Maxted}}]{2020AJ....159....4W}
{Wang}, L., {Gies}, D.~R., {Lester}, K.~V., {et~al.} 2020{\natexlab{b}}, \aj, 159, 4, \dodoi{10.3847/1538-3881/ab52fa}

\bibitem[{{Wright} {et~al.}(2010){Wright}, {Eisenhardt}, {Mainzer}, {Ressler}, {Cutri}, {Jarrett}, {Kirkpatrick}, {Padgett}, {McMillan}, {Skrutskie}, {Stanford}, {Cohen}, {Walker}, {Mather}, {Leisawitz}, {Gautier}, {McLean}, {Benford}, {Lonsdale}, {Blain}, {Mendez}, {Irace}, {Duval}, {Liu}, {Royer}, {Heinrichsen}, {Howard}, {Shannon}, {Kendall}, {Walsh}, {Larsen}, {Cardon}, {Schick}, {Schwalm}, {Abid}, {Fabinsky}, {Naes}, \& {Tsai}}]{2010AJ....140.1868W}
{Wright}, E.~L., {Eisenhardt}, P. R.~M., {Mainzer}, A.~K., {et~al.} 2010, \aj, 140, 1868, \dodoi{10.1088/0004-6256/140/6/1868}

\bibitem[{{Xiong} {et~al.}(2024){Xiong}, {Ding}, {Li}, {Ge}, {Cheng}, {Ji}, {Han}, \& {Chen}}]{2024ApJS..270...20X}
{Xiong}, J., {Ding}, X., {Li}, J., {et~al.} 2024, \apjs, 270, 20, \dodoi{10.3847/1538-4365/ad0ceb}

\bibitem[{{York} {et~al.}(2000){York}, {Adelman}, {Anderson}, {Anderson}, {Annis}, {Bahcall}, {Bakken}, {Barkhouser}, {Bastian}, {Berman}, {Boroski}, {Bracker}, {Briegel}, {Briggs}, {Brinkmann}, {Brunner}, {Burles}, {Carey}, {Carr}, {Castander}, {Chen}, {Colestock}, {Connolly}, {Crocker}, {Csabai}, {Czarapata}, {Davis}, {Doi}, {Dombeck}, {Eisenstein}, {Ellman}, {Elms}, {Evans}, {Fan}, {Federwitz}, {Fiscelli}, {Friedman}, {Frieman}, {Fukugita}, {Gillespie}, {Gunn}, {Gurbani}, {de Haas}, {Haldeman}, {Harris}, {Hayes}, {Heckman}, {Hennessy}, {Hindsley}, {Holm}, {Holmgren}, {Huang}, {Hull}, {Husby}, {Ichikawa}, {Ichikawa}, {Ivezi{\'c}}, {Kent}, {Kim}, {Kinney}, {Klaene}, {Kleinman}, {Kleinman}, {Knapp}, {Korienek}, {Kron}, {Kunszt}, {Lamb}, {Lee}, {Leger}, {Limmongkol}, {Lindenmeyer}, {Long}, {Loomis}, {Loveday}, {Lucinio}, {Lupton}, {MacKinnon}, {Mannery}, {Mantsch}, {Margon}, {McGehee}, {McKay}, {Meiksin}, {Merelli}, {Monet}, {Munn}, {Narayanan}, {Nash}, {Neilsen}, {Neswold}, {Newberg}, {Nichol}, {Nicinski},
  {Nonino}, {Okada}, {Okamura}, {Ostriker}, {Owen}, {Pauls}, {Peoples}, {Peterson}, {Petravick}, {Pier}, {Pope}, {Pordes}, {Prosapio}, {Rechenmacher}, {Quinn}, {Richards}, {Richmond}, {Rivetta}, {Rockosi}, {Ruthmansdorfer}, {Sandford}, {Schlegel}, {Schneider}, {Sekiguchi}, {Sergey}, {Shimasaku}, {Siegmund}, {Smee}, {Smith}, {Snedden}, {Stone}, {Stoughton}, {Strauss}, {Stubbs}, {SubbaRao}, {Szalay}, {Szapudi}, {Szokoly}, {Thakar}, {Tremonti}, {Tucker}, {Uomoto}, {Vanden Berk}, {Vogeley}, {Waddell}, {Wang}, {Watanabe}, {Weinberg}, {Yanny}, {Yasuda}, \& {SDSS Collaboration}}]{2000AJ....120.1579Y}
{York}, D.~G., {Adelman}, J., {Anderson}, John~E., J., {et~al.} 2000, \aj, 120, 1579, \dodoi{10.1086/301513}

\bibitem[{{Zhang} {et~al.}(2023){Zhang}, {Green}, \& {Rix}}]{2023MNRAS.524.1855Z}
{Zhang}, X., {Green}, G.~M., \& {Rix}, H.-W. 2023, \mnras, 524, 1855, \dodoi{10.1093/mnras/stad1941}

\bibitem[{{Zhang} {et~al.}(2017){Zhang}, {Fu}, {Liu}, {Luo}, \& {Ren}}]{2017ApJ...850..125Z}
{Zhang}, X.~B., {Fu}, J.~N., {Liu}, N., {Luo}, C.~Q., \& {Ren}, A.~B. 2017, \apj, 850, 125, \dodoi{10.3847/1538-4357/aa9577}

\bibitem[{{Zhang} {et~al.}(2013){Zhang}, {Luo}, \& {Fu}}]{2013ApJ...777...77Z}
{Zhang}, X.~B., {Luo}, C.~Q., \& {Fu}, J.~N. 2013, \apj, 777, 77, \dodoi{10.1088/0004-637X/777/1/77}

\bibitem[{{Zhang} {et~al.}(2019){Zhang}, {Wang}, {Chen}, {Luo}, \& {Zhang}}]{2019ApJ...884..165Z}
{Zhang}, X.~B., {Wang}, K., {Chen}, X.~H., {Luo}, C.~Q., \& {Zhang}, C.~G. 2019, \apj, 884, 165, \dodoi{10.3847/1538-4357/ab3fa9}

\bibitem[{{Zhao} {et~al.}(2012){Zhao}, {Zhao}, {Chu}, {Jing}, \& {Deng}}]{2012RAA....12..723Z}
{Zhao}, G., {Zhao}, Y.-H., {Chu}, Y.-Q., {Jing}, Y.-P., \& {Deng}, L.-C. 2012, Research in Astronomy and Astrophysics, 12, 723, \dodoi{10.1088/1674-4527/12/7/002}

\end{thebibliography}
\bibliographystyle{aasjournal}

\begin{appendices}

\section{Appendix A}\label{AppendixA}
\begin{table*}[htbp]
\tiny 
\caption{Absolute parameters of 29 EL CVn-type binaries from TESS survey.}
\centering
\begin{tabular}{lccccccccccc}
\hline
\hline
  \multicolumn{1}{l}{TIC} &
  \multicolumn{1}{c}{Period (d)} &
  \multicolumn{1}{c}{$T\mathrm{_{1}}$ (K)} &
  \multicolumn{1}{c}{$T\mathrm{_{2}}$ (K)} &
  \multicolumn{1}{c}{\logg$_{1}$ [cgs]} &
  \multicolumn{1}{c}{\logg$_{2}$ [cgs]} &
    \multicolumn{1}{c}{$M\mathrm{_{1}}$ ($M_{\odot}$)} &
  \multicolumn{1}{c}{$M\mathrm{_{2}}$ ($M_{\odot}$)}&
  \multicolumn{1}{c}{$R\mathrm{_{1}}$ ($R_{\odot}$)} &
  \multicolumn{1}{c}{$R\mathrm{_{2}}$ ($R_{\odot}$)}\\
\hline
\hline
$^{1}$35399970 & 1.2908616 & 8597 $^{+ 68 }_{- 39 }$& 11378 $^{+ 83 }_{- 82 }$& 4.294 $^{+ 0.003  }_{- 0.003  }$& 4.971 $^{+ 0.005 }_{- 0.005  }$& 1.850  $^{+ 0.020  }_{- 0.020  }$& 0.193  $^{+ 0.002  }_{- 0.002  }$& 1.604  $^{+ 0.007  }_{- 0.007  }$& 0.238  $^{+ 0.001  }_{- 0.001}$\\
$^{1}$83833793 & 2.17351686 & 7903 $^{+ 6 }_{- 49 }$& 10044 $^{+ 37 }_{- 35 }$& 3.871 $^{+ 0.004  }_{- 0.004  }$& 4.185 $^{+ 0.072  }_{- 0.072  }$& 2.050  $^{+ 0.020  }_{- 0.020  }$& 0.200  $^{+ 0.030  }_{- 0.030  }$& 2.749  $^{+ 0.016  }_{- 0.016  }$& 0.598  $^{+ 0.006  }_{- 0.006}$\\
$^{1}$121078334 & 0.92859504 & 7644 $^{+ 27 }_{- 6 }$& 10844 $^{+ 22 }_{- 22 }$& 4.143 $^{+ 0.002  }_{- 0.002  }$& 4.741 $^{+ 0.002  }_{- 0.002  }$& 1.680  $^{+ 0.020  }_{- 0.010  }$& 0.185  $^{+ 0.002  }_{- 0.002  }$& 1.818  $^{+ 0.006  }_{- 0.006  }$& 0.303  $^{+ 0.001  }_{- 0.001}$\\
$^{1}$160081043 & 0.768701771 & 8020 $^{+ 3 }_{- 47 }$& 9985 $^{+ 30 }_{- 29 }$& 4.293 $^{+ 0.001  }_{- 0.001  }$& 4.427 $^{+ 0.001  }_{- 0.001  }$& 1.690  $^{+ 0.010  }_{- 0.010  }$& 0.196  $^{+ 0.001  }_{- 0.001  }$& 1.537  $^{+ 0.003  }_{- 0.003  }$& 0.447  $^{+ 0.001  }_{- 0.001}$\\
$^{1}$35399970 & 1.2908616 & 8597 $^{+ 68 }_{- 39 }$& 11378 $^{+ 83 }_{- 82 }$& 4.294 $^{+ 0.003  }_{- 0.003  }$& 4.971 $^{+ 0.005 }_{- 0.005  }$& 1.850  $^{+ 0.020  }_{- 0.020  }$& 0.193  $^{+ 0.002  }_{- 0.002  }$& 1.604  $^{+ 0.007  }_{- 0.007  }$& 0.238  $^{+ 0.001  }_{- 0.001}$\\
$^{1}$166874908 & 2.18931 & 7928 $^{+ 4 }_{- 30 }$& 11127 $^{+ 24 }_{- 22 }$&3.954 $^{+ 0.002  }_{- 0.002  }$& 4.702 $^{+ 0.010  }_{- 0.010  }$& 1.950  $^{+ 0.010  }_{- 0.010  }$& 0.197  $^{+ 0.004  }_{- 0.004  }$& 2.436  $^{+ 0.006  }_{- 0.006  }$& 0.327  $^{+ 0.002  }_{- 0.002}$\\
$^{1}$192990023 & 0.900898203 & 7198 $^{+ 8 }_{- 47 }$& 9288 $^{+ 34 }_{- 35 }$& 4.169 $^{+ 0.004  }_{- 0.004  }$& 4.633 $^{+ 0.005  }_{- 0.005  }$& 1.530  $^{+ 0.040  }_{- 0.040  }$& 0.193  $^{+ 0.005  }_{- 0.005  }$& 1.685  $^{+ 0.016  }_{- 0.016  }$& 0.350  $^{+ 0.003  }_{- 0.003}$\\
$^{1}$408351887 & 1.0731802 & 7113 $^{+ 2 }_{- 2 }$& 10263 $^{+ 4 }_{- 4 }$& 3.909 $^{+ 0.007}_{- 0.007}$& 4.633 $^{+ 0.017  }_{- 0.017  }$& 1.740  $^{+ 0.010  }_{- 0.010  }$& 0.196  $^{+ 0.005  }_{- 0.005}$& 2.425  $^{+ 0.021  }_{- 0.021  }$& 0.353  $^{+ 0.005  }_{- 0.005}$\\
$^{1}$149160359 & 1.120738 & 7827 $^{+ 22 }_{- 61 }$& 8583 $^{+ 49 }_{- 49 }$& 4.009 $^{+ 0.001  }_{- 0.001  }$& 4.306 $^{+ 0.001  }_{- 0.001  }$& 1.880  $^{+ 0.001  }_{- 0.001  }$& 0.188  $^{+ 0.001  }_{- 0.001  }$& 2.245  $^{+ 0.001  }_{- 0.001  }$& 0.505  $^{+ 0.001  }_{- 0.001}$\\
$^{1}$416264037 & 1.15991 & 7101 $^{+ 15 }_{- 1 }$& 9649 $^{+ 13 }_{- 14 }$& 3.983 $^{+ 0.001}_{- 0.001}$& 4.517 $^{+ 0.001  }_{- 0.001  }$& 1.660  $^{+ 0.001  }_{- 0.001  }$& 0.178  $^{+ 0.001  }_{- 0.001}$& 2.175  $^{+ 0.001  }_{- 0.001  }$& 0.385  $^{+ 0.001  }_{- 0.001}$\\
$^{1}$100011519 & 1.735248 & 7573 $^{+ 2}_{- 9 }$& 8447 $^{+ 6 }_{- 6 }$& 3.857 $^{+ 0.001  }_{- 0.001  }$& 4.183 $^{+ 0.001  }_{- 0.001  }$& 1.960  $^{+ 0.001  }_{- 0.001  }$& 0.204  $^{+ 0.001  }_{- 0.001  }$& 2.733  $^{+ 0.001  }_{- 0.001  }$& 0.606  $^{+ 0.001  }_{- 0.001}$\\
$^{1}$219485855 & 0.660002 & 7808 $^{+ 69 }_{- 2 }$& 7903 $^{+ 52 }_{- 48 }$& 4.149 $^{+ 0.001}_{- 0.001}$& 4.217 $^{+ 0.001  }_{- 0.001}$& 1.740  $^{+ 0.001  }_{- 0.001 }$& 0.183  $^{+ 0.001  }_{- 0.001}$& 1.838  $^{+ 0.001  }_{- 0.001  }$& 0.552  $^{+ 0.001  }_{- 0.001}$\\
$^{1}$399725538 & 1.293273 & 7761 $^{+ 66 }_{- 54 }$& 11555 $^{+ 99 }_{- 99 }$& 4.144 $^{+ 0.002  }_{- 0.002  }$& 5.131 $^{+ 0.002  }_{- 0.002  }$& 1.720  $^{+ 0.030  }_{- 0.010  }$& 0.189  $^{+ 0.002  }_{- 0.002  }$& 1.839 $^{+ 0.007  }_{- 0.007}$& 0.196  $^{+ 0.001  }_{- 0.001}$\\
$^{1}$464641792 & 1.369715 & 7499 $^{+ 1 }_{- 14 }$& 9750 $^{+ 10 }_{- 11 }$& 3.937 $^{+ 0.001}_{- 0.001}$& 4.583 $^{+ 0.001  }_{- 0.001  }$& 1.840  $^{+ 0.010  }_{- 0.010 }$& 0.184  $^{+ 0.001  }_{- 0.001}$& 2.415  $^{+ 0.005  }_{- 0.005  }$& 0.363  $^{+ 0.001  }_{- 0.001}$\\

$^{2}$54957535 & 0.792833 & 6650  $_{- 2  }^{+ 17  }$& 9371  $_{- 17  }^{+ 16  }$& 4.041 $_{- 0.002 }^{+ 0.002 }$& 4.706 $_{- 0.002 }^{+ 0.002 }$& 1.460 $_{- 0.020 }^{+ 0.020 }$& 0.171 $_{- 0.002 }^{+ 0.002 }$& 1.908 $_{- 0.009 }^{+ 0.009 }$& 0.303 $_{- 0.001 }^{+ 0.001 }$\\
$^{2}$142258314 & 0.8441541 & 8304  $_{- 37  }^{+ 84  }$& 10875  $_{- 92  }^{+ 89  }$& 4.214 $_{- 0.010 }^{+ 0.100 }$& 4.952 $_{- 0.017 }^{+ 0.017 }$& 1.830 $_{- 0.030 }^{+ 0.100 }$& 0.179 $_{- 0.007 }^{+ 0.007 }$& 1.749 $_{- 0.027 }^{+ 0.028 }$& 0.234 $_{- 0.005 }^{+ 0.005 }$\\
$^{2}$197604137 & 1.16189659 & 7044  $_{- 28  }^{+ 35  }$& 9731  $_{- 46  }^{+ 47  }$& 3.936 $_{- 0.016 }^{+ 0.014 }$& 4.846 $_{- 0.034 }^{+ 0.034 }$& 1.690 $_{- 0.020 }^{+ 0.020 }$& 0.189 $_{- 0.009 }^{+ 0.009 }$& 2.314 $_{- 0.040 }^{+ 0.040 }$& 0.272 $_{- 0.008 }^{+ 0.008 }$\\
$^{2}$400028476 & 0.64410724 & 7297  $_{- 11  }^{+ 3  }$& 9432  $_{- 9  }^{+ 9  }$& 4.145 $_{- 0.001 }^{+ 0.001 }$& 4.588 $_{- 0.001 }^{+ 0.001 }$& 1.550 $_{- 0.001 }^{+ 0.001 }$& 0.188 $_{- 0.001 }^{+ 0.001 }$& 1.744 $_{- 0.001 }^{+ 0.001 }$& 0.364 $_{- 0.001 }^{+ 0.001 }$\\
$^{2}$420947520 & 1.799427277 & 8291  $_{- 27  }^{+ 5  }$& 11597  $_{- 25  }^{+ 22  }$& 4.065 $_{- 0.006 }^{+ 0.006 }$& 5.039 $_{- 0.016 }^{+ 0.015 }$& 1.950 $_{- 0.030 }^{+ 0.030 }$& 0.189 $_{- 0.006 }^{+ 0.006 }$& 2.144 $_{- 0.018 }^{+ 0.018 }$& 0.218 $_{- 0.002 }^{+ 0.002 }$\\
$^{2}$464641792 & 1.369715 & 7499  $_{- 14  }^{+ 0  }$& 9750  $_{- 12  }^{+ 11  }$& 3.936 $_{- 0.001 }^{+ 0.001 }$& 4.583 $_{- 0.001 }^{+ 0.001 }$& 1.840 $_{- 0.010 }^{+ 0.010 }$& 0.184 $_{- 0.001 }^{+ 0.001 }$& 2.415 $_{- 0.005 }^{+ 0.005 }$& 0.363 $_{- 0.001 }^{+ 0.001 }$\\
$^{3}$28763683 & 0.892 & 8756  $_{- 1  }^{+ 5  }$& 9303  $_{- 4  }^{+ 3  }$& 3.998 $_{- 0.003 }^{+ 0.003 }$& 4.297 $_{- 0.004 }^{+ 0.004 }$& 2.210 $_{- 0.090 }^{+ 0.001 }$& 0.196 $_{- 0.005 }^{+ 0.005 }$& 2.465 $_{- 0.019 }^{+ 0.019 }$& 0.522 $_{- 0.004 }^{+ 0.004 }$\\
$^{3}$53896096 & 0.741 & 7923  $_{- 7  }^{+ 6  }$& 10578  $_{- 44  }^{+ 46  }$& 4.240 $_{- 0.025 }^{+ 0.023 }$& 5.098 $_{- 0.076 }^{+ 0.075 }$& 1.690 $_{- 0.010 }^{+ 0.020 }$& 0.171 $_{- 0.019 }^{+ 0.019 }$& 1.633 $_{- 0.043 }^{+ 0.048 }$& 0.192 $_{- 0.011 }^{+ 0.011 }$\\
$^{3}$65448527 & 0.66783 & 7843  $_{- 6  }^{+ 13  }$& 9805  $_{- 13  }^{+ 12  }$& 4.281 $_{- 0.001 }^{+ 0.001 }$& 4.566 $_{- 0.002 }^{+ 0.002 }$& 1.650 $_{- 0.020 }^{+ 0.020 }$& 0.168 $_{- 0.001 }^{+ 0.001 }$& 1.537 $_{- 0.003 }^{+ 0.003 }$& 0.354 $_{- 0.001 }^{+ 0.001 }$\\
$^{3}$165371937 & 0.79564 & 8097  $_{- 59  }^{+ 42  }$& 10598  $_{- 84  }^{+ 85  }$& 4.298 $_{- 0.004 }^{+ 0.004 }$& 4.796 $_{- 0.004 }^{+ 0.004 }$& 1.640 $_{- 0.010 }^{+ 0.080 }$& 0.188 $_{- 0.005 }^{+ 0.005 }$& 1.503 $_{- 0.015 }^{+ 0.014 }$& 0.287 $_{- 0.003 }^{+ 0.003 }$\\
$^{3}$430145146 & 0.744 & 8053  $_{- 9  }^{+ 18  }$& 10844  $_{- 21  }^{+ 21  }$& 4.222 $_{- 0.003 }^{+ 0.003 }$& 4.831 $_{- 0.005 }^{+ 0.005 }$& 1.740 $_{- 0.010 }^{+ 0.010 }$& 0.188 $_{- 0.002 }^{+ 0.002 }$& 1.691 $_{- 0.006 }^{+ 0.006 }$& 0.276 $_{- 0.001 }^{+ 0.001 }$\\
$^{3}$471013508 & 1.163 & 7470  $_{- 2  }^{+ 4  }$& 8824  $_{- 5  }^{+ 5  }$& 3.893 $_{- 0.002 }^{+ 0.002 }$& 3.627 $_{- 0.006 }^{+ 0.006 }$& 1.870 $_{- 0.010 }^{+ 0.010 }$& 0.239 $_{- 0.003 }^{+ 0.003 }$& 2.560 $_{- 0.007 }^{+ 0.007 }$& 1.245 $_{- 0.004 }^{+ 0.004 }$\\
$^{3}$85277226 & 0.977 & 7984  $_{- 37  }^{+ 25  }$& 8404  $_{- 110  }^{+ 102  }$& 4.289 $_{- 0.005 }^{+ 0.005 }$& 4.435 $_{- 0.042 }^{+ 0.040 }$& 1.680 $_{- 0.010 }^{+ 0.010 }$& 0.146 $_{- 0.013 }^{+ 0.013 }$& 1.538 $_{- 0.009 }^{+ 0.010 }$& 0.383 $_{- 0.004 }^{+ 0.004 }$\\
$^{3}$110981024 & 1.9793094 & 8398  $_{- 143  }^{+ 15  }$& 16557  $_{- 159  }^{+ 153  }$& 3.692 $_{- 0.013 }^{+ 0.013 }$& 5.037 $_{- 0.028 }^{+ 0.028 }$& 2.330 $_{- 0.090 }^{+ 0.090 }$& 0.214 $_{- 0.012 }^{+ 0.013 }$& 3.597 $_{- 0.070 }^{+ 0.071 }$& 0.232 $_{- 0.006 }^{+ 0.006 }$\\
$^{3}$177118177 & 2.5046225 & 8468  $_{- 6  }^{+ 10  }$& 14661  $_{- 15  }^{+ 15  }$& 3.916 $_{- 0.038 }^{+ 0.035 }$& 5.666 $_{- 0.093 }^{+ 0.085 }$& 2.170 $_{- 0.040 }^{+ 0.050 }$& 0.201 $_{- 0.032 }^{+ 0.032 }$& 2.685 $_{- 0.104 }^{+ 0.124 }$& 0.108 $_{- 0.006 }^{+ 0.006 }$\\
$^{3}$234874474 & 1.3159894 & 10647  $_{- 468  }^{+ 68  }$& 19846  $_{- 451  }^{+ 432  }$& 4.071 $_{- 0.006 }^{+ 0.006 }$& 5.361 $_{- 0.012 }^{+ 0.012 }$& 2.490 $_{- 0.010 }^{+ 0.02 }$& 0.228 $_{- 0.002 }^{+ 0.002 }$& 2.409 $_{- 0.017 }^{+ 0.018 }$& 0.165 $_{- 0.002 }^{+ 0.002 }$\\
$^{3}$424008045 & 1.1836627 & 7273  $_{- 3  }^{+ 2  }$& 9279  $_{- 4  }^{+ 4  }$& 3.900 $_{- 0.004 }^{+ 0.004 }$& 4.342 $_{- 0.013 }^{+ 0.013 }$& 1.800 $_{- 0.010}^{+ 0.010 }$& 0.179 $_{- 0.004 }^{+ 0.004 }$& 2.491 $_{- 0.011 }^{+ 0.012 }$& 0.473 $_{- 0.005 }^{+ 0.005 }$\\
\hline
\multicolumn{5}{l}{\footnotesize $^{1}$ previously discovered with parameters determined}\\
\multicolumn{5}{l}{\footnotesize $^{2}$ previously discovered}\\
\multicolumn{5}{l}{\footnotesize $^{3}$ new discovered}\\
\end{tabular}
\end{table*}

\end{appendices}


\end{CJK}
\end{document}